\pdfoutput=1

\documentclass[camera,letterpaper,nomarginnotes,nonarrowgutter]{jpaper}

\usepackage[normalem]{ulem}
\usepackage{subcaption}
\usepackage{clipboard}
\usepackage{soul}
\usepackage{tikz}
\usepackage{flushend}
\usepackage[noadjust]{cite}
\usepackage{xargs} 
\usepackage{mathtools}
\usepackage{enumitem}
\usepackage{booktabs}
\usepackage{multirow}
\usepackage[many]{tcolorbox}% version 2.80 (2014/03/31)
\usepackage{longfbox}

\usepackage{algorithmic}
\usepackage{graphicx}
\usepackage{textcomp}
\usepackage{xcolor}
\usepackage{pifont}
\usepackage[acronym,nonumberlist,nowarn]{glossaries}
    \glsdisablehyper
     \newacronym{ADI}{ADI}{Alternating Direction Implicit}
\newacronym{AGU}{AGU}{Address Generation Unit}
\newacronym[longplural={Access History Tables}]{AHT}{AHT}{Access History Table}
\newacronym{AHT-R}{AHT-R}{Access History Table - Read}
\newacronym{AHT-W}{AHT-W}{Access History Table - Write-Back}
\newacronym{AIP}{AIP}{Access Interval Predictor}
\newacronym{AL}{AL}{Added Latency for column accesses}
\newacronym{ASIC}{ASIC}{Application Specific Integrated Circuit}
\newacronym{AVX}{AVX}{Advanced Vector Extensions}
\newacronym[longplural={Basic Block Vectors}]{BBV}{BBV}{Basic Block Vector}
\newacronym{BHT}{BHT}{Branch History Table}
\newacronym{HBM}{HBM}{Hybrid Bandwidth Memory}
\newacronym{DDR4}{DDR4}{Double-Data Rate 4}
\newacronym{MIMD}{MIMD}{multiple-instruction multiple-data}
\newacronym{SIMT}{SIMT}{Single Instruction Multiple Threads}
\newacronym{SLP}{SALP}{subarray-level parallelism}
\newacronym{BLP}{BLP}{bank-level parallelism}

\newacronym{NN}{NN}{neural network}
\newacronym{BTB}{BTB}{Branch Target Buffer}
\newacronym{CAS}{CAS}{Column Address Strobe}
\newacronym{AMAT}{AMAT}{Average Memory Access Time}
\newacronym{CCD}{CCD}{Column to Column Delay}
\newacronym{AI}{AI}{Arithmetic Intensity}
\newacronym[longplural={Columnar Database Systems}]{CDBS}{CDBS}{Columnar Database System}
\newacronym[longplural={Chip Multiprocessors}]{CMP}{CMP}{Chip Multiprocessor}
\newacronym{CMT}{CMT}{Chip Multithreading}
\newacronym[longplural={Coarse-Grain Reconfigurable Arrays}]{CGRA}{CGRA}{Coarse-Grain Reconfigurable Array}
\newacronym{C-RAM}{C-RAM}{Computational-RAM}
\newacronym{CWD}{CWD}{Column Write Delay}
\newacronym{DBI}{DBI}{Dynamic Binary Instrumentation}
\newacronym[longplural={Database Management Systems}]{DBMS}{DBMS}{Database Management System}
\newacronym{DDR}{DDR}{Double Data Rate}
\newacronym{DEWP}{DEWP}{Dead Line and Early Write-Back Predictor}
\newacronym{DRAM}{DRAM}{Dynamic Random Access Memory}
\newacronym{DSBP}{DSBP}{Dead Sub-Block Predictor}
\newacronym{DSM}{DSM}{Decomposed Storage Manage}
\newacronym{FAW}{FAW}{Four row Activation Window}
\newacronym{FP}{FP}{Floating-Point}
\newacronym{EDP}{EDP}{Energy-Delay Product}
\newacronym{FCFS}{FCFS}{First-Come First-Serve}
\newacronym{FIFO}{FIFO}{Fist In, First Out}
\newacronym{FSM}{FSM}{finite state machine}
\newacronym{FPGA}{FPGA}{Field-Programmable Gate Array}
\newacronym[longplural={Functional Units}]{FU}{FU}{Functional Unit}
\newacronym{GAg}{GAg}{Global Adaptive branch prediction using one Global PHT}
\newacronym{GAs}{GAs}{Global Adaptive branch prediction using per-Set PHT}
\newacronym{GCC}{GCC}{GNU Compiler Collection}
\newacronym{GEMS}{GEMS}{General Execution-driven Multiprocessor Simulator}
\newacronym[longplural={General Purpose Processors}]{GPP}{GPP}{General Purpose Processor}
\newacronym{HMC}{HMC}{Hybrid Memory Cube}
\newacronym{HIVE}{HIVE}{HMC Instruction Vector Extensions}
\newacronym{IATAC}{IATAC}{Inter-Access Time per Access Count}
\newacronym{ILP}{ILP}{Instruction Level Parallelism}
\newacronym{IPC}{IPC}{Instructions per Cycle}
\newacronym{ISA}{ISA}{Instruction Set Architecture}
\newacronym{IRAM}{IRAM}{Intelligent RAM}
\newacronym{KIPS}{KIPS}{Kilo Instructions per Second}
\newacronym{LDS}{LDS}{Linked Data Structure}
\newacronym{LFSR}{LFSR}{Linear Feedback Shift Register}
\newacronym{LFMR}{LFMR}{Last-to-First Miss-Ratio}
\newacronym{MLP}{MLP}{Memory-Level Parallelism}

\newacronym{LLC}{LLC}{last-level cache}
\newacronym{LRU}{LRU}{Least Recently Used}
\newacronym{LSU}{LSU}{Load Store Unit}
\newacronym{LTP}{LTP}{Last-Touch Predictor}
\newacronym{LvP}{LvP}{Live-time Predictor}
\newacronym{LWP}{LWP}{Last Write Predictor}
\newacronym{MARSS}{MARSS}{Micro Architectural and System Simulator}
\newacronym{McPAT}{McPAT}{Multi-core Power, Area, and Timing}
\newacronym{MIPS}{MIPS}{Microprocessor without Interlocked Pipeline Stages}
\newacronym{MOB}{MOB}{Memory Order Buffer}
\newacronym{MMU}{MMU}{memory management unit}
\newacronym{PuD}{PUD}{processing-using-DRAM}
\newacronym{MPKI}{MPKI}{Misses per Kilo-Instruction}
\newacronym{MSHR}{MSHR}{Miss-Status Handling Registers}
\newacronym{NAS}{NAS}{Numerical Aerodynamic Simulation}
\newacronym{NDP}{NDP}{Near-Data Processing}
\newacronym{NMP}{NMP}{Near Memory Processor}
\newacronym{NoC}{NoC}{Network-on-Chip}
\newacronym{NPB}{NPB}{NAS Parallel Benchmark}
\newacronym{NUCA}{NUCA}{Non-Uniform Cache Architecture}
\newacronym{NUMA}{NUMA}{Non-Uniform Memory Access}
\newacronym{OoO}{OoO}{Out-of-Order}
\newacronym{OpenMP}{OpenMP}{Open Multi-Processing}
\newacronym{OS}{OS}{Operating System}
\newacronym{PAg}{PAg}{Per-address Adaptive branch prediction using one Global PHT}
\newacronym{PAs}{PAs}{Per-address Adaptive branch prediction using per-Set PHT}
\newacronym{PC}{PC}{Program Counter}
\newacronym[longplural={Pointer-Chasing Engines}]{PCE}{PCE}{Pointer-Chasing Engine}
\newacronym{PCM}{PCM}{Performance Counter Monitor}
\newacronym{PIM}{PIM}{processing-in-memory}
\newacronym[longplural={Pattern History Tables}]{PHT}{PHT}{Pattern History Table}
\newacronym{RAPL}{RAPL}{Running Average Power Limit}
\newacronym{GEMM}{GEMM}{general matrix-matrix multiplication}
\newacronym{RAS}{RAS}{Row Address Strobe}
\newacronym{RAT}{RAT}{Registers Alias Table}
\newacronym{RC}{RC}{Row Cycle}
\newacronym{RCD}{RCD}{RAS to CAS Delay}
\newacronym[longplural={Row-based Database Systems}]{RDBS}{RDBS}{Row-based Database System}
\newacronym{ROB}{ROB}{Reorder Buffer}
\newacronym{RRD}{RRD}{Row to Row activation Delay}
\newacronym{DNN}{DNN}{Deep Neural Network}
\newacronym{ANN}{ANN}{Artificial Neural Network}
\newacronym{FFD}{FFD}{first-fit-decreasing}
\newacronym{HFF}{HFF}{helper flip-flop}
\newacronym{OBPS}{OBPS}{one-bit per-subarray}
\newacronym{ABOS}{ABOS}{all-bits in one-subarray}
\newacronym{ABPS}{ABPS}{all-bits per-subarray}
\newacronym{GEMV}{GEMV}{general matrix-vector product}

\newacronym{RBR}{RBR}{redundant binary representation}
\newacronym{RP}{RP}{Row Precharge}
\newacronym{RTL}{RTL}{Register Transfer Level}
\newacronym{RTP}{RTP}{Read To Precharge}
\newacronym{RVU}{RVU}{Reconfigurable Vector Unit}
\newacronym{SDP}{SDP}{Skewed Dead-Block Predictor}
\newacronym{SESC}{SESC}{Superescalar Simulator}
\newacronym{SSE}{SSE}{Streaming SIMD Extensions}
\newacronym{SFP}{SFP}{Spatial Footprint Predictor}
\newacronym{SiNUCA}{SiNUCA}{Simulator of Non-Uniform Cache Architectures}
\newacronym{SMT}{SMT}{Simultaneous Multi-Threading}
\newacronym{SIMD}{SIMD}{single-instruction multiple-data}
\newacronym{LUT}{LUT}{Lookup Table}

\newacronym{SPEC}{SPEC}{Standard Performance Evaluation Corporation}
\newacronym{SoC}{SoC}{System-on-Chip}
\newacronym{SPP}{SPP}{Spatial Pattern Predictor}
\newacronym{SRAM}{SRAM}{Static Random Access Memory}
\newacronym{SSOR}{SSOR}{Symmetric Successive Over-Relaxation}
\newacronym{SSV}{SSV}{Search Set Vector}
\newacronym{TLB}{TLB}{Translation Look-aside Buffer}
\newacronym{TLP}{TLP}{Thread Level Parallelism}
\newacronym[longplural={Through-Silicon Vias}]{TSV}{TSV}{Through-Silicon Via}
\newacronym{VWQ}{VWQ}{Virtual Write Queue}
\newacronym{WTR}{WTR}{Write Recovery time}
\newacronym{WR}{WR}{Write To Read delay time}
\newacronym{TRA}{TRA}{triple row activation}

\usepackage{xspace}
\usepackage[binary-units=true]{siunitx}
\usepackage[us,12hr]{datetime}
\usepackage[en-GB, useregional=numeric]{datetime2}
\usepackage{balance}
\usepackage{dblfloatfix}

\usepackage{setspace}
\usepackage[export]{adjustbox}

\usepackage{fancyhdr}

\usepackage{cleveref}
\crefformat{section}{\S#2#1#3}
\crefformat{subsection}{\S#2#1#3}
\crefformat{subsubsection}{\S#2#1#3}

\definecolor{airforceblue}{rgb}{0.36, 0.54, 0.66}
\definecolor{dodgerblue}{rgb}{0.12, 0.56, 1.0}
\definecolor{brandeisblue}{rgb}{0.0, 0.44, 1.0}
\definecolor{brickred}{rgb}{0.8, 0.25, 0.33}
\definecolor{eggplant}{rgb}{0.38, 0.25, 0.32}
\definecolor{byzantium}{rgb}{0.44, 0.16, 0.39}
\definecolor{ddgreen}{rgb}{0.00, 0.50, 0.00}

\definecolor{mygreen}{rgb}{0,0.6,0}
\definecolor{mygray}{rgb}{0.5,0.5,0.5}
\definecolor{mymauve}{rgb}{0.58,0,0.82}

\definecolor{bluehl}{rgb}{0.8,0.874,1}
\definecolor{pinkhl}{rgb}{0.992156863,0.847058824,1}
\definecolor{macaroniandcheese}{rgb}{1.0, 0.74, 0.53}
\definecolor{mossgreen}{rgb}{0.68, 0.87, 0.68}
\definecolor{greenhl}{rgb}{0.835,0.996,0.939}
\definecolor{yellowhl}{rgb}{0.996,0.957,0.8}
\definecolor{palecerulean}{rgb}{0.61, 0.77, 0.89}
\definecolor{gray(x11gray)}{rgb}{0.75, 0.75, 0.75}
\definecolor{amethyst}{rgb}{0.6, 0.4, 0.8}
\definecolor{ao}{rgb}{0.0, 0.5, 0.0}
\definecolor{burntorange}{rgb}{0.8, 0.33, 0.0}

\definecolor{cadmiumorange}{rgb}{0.93, 0.53, 0.18}

\definecolor{frenchlilac}{rgb}{0.53, 0.38, 0.56}
\definecolor{heliotrope}{rgb}{0.87, 0.45, 1.0}
\definecolor{peridot}{rgb}{0.9, 0.89, 0.0}
\definecolor{saffron}{rgb}{0.96, 0.77, 0.19}
\definecolor{tuscanred}{rgb}{0.51, 0.21, 0.21}
\definecolor{uscgold}{rgb}{1.0, 0.8, 0.0}
\definecolor{tangerineyellow}{rgb}{1.0, 0.8, 0.0}
\definecolor{rufous}{rgb}{0.66, 0.11, 0.03}
\definecolor{safetyorange}{rgb}{1.0, 0.4, 0.0}

\newif\ifcameraready
\camerareadytrue
\ifcameraready

    \newcommand{\gfcrii}[1]{\textcolor{black}{#1}}

\else

    \newcommand{\gfcrii}[1]{\textcolor{black}{#1}}

\fi

\newcommand{\versionnum}[0]{4.0}

\definecolor{MidnightBlue}{rgb}{0.1, 0.1, 0.44}

\definecolor{blush}{rgb}{0.87, 0.36, 0.51}

\newcommand\ignore[1]{ }
\newcommand{\revdel}[1]{}
\newcommand{\sgdel}[1]{}

\newif\ifasplosrevisionsubmission
\asplosrevisionsubmissiontrue
\ifasplosrevisionsubmission
    
\else
    
\fi

\newif\ifasplosrevision
\asplosrevisionfalse
\ifasplosrevision 
    
\else

\fi

\newif\ifasplossubmission
\asplossubmissiontrue
\ifasplossubmission

    \newcommand{\agyasploscomment}[1]{}
\else

    \newcommand{\agyasploscomment}[1]{\textcolor{red}{\textbf{!!!~Giray:} #1}}
    
\fi

\newif\ifmicrosubmission
\microsubmissiontrue
\ifmicrosubmission

    \newcommand{\agymicrocomment}[1]{}
\else

    \newcommand{\agymicrocomment}[1]{\textcolor{red}{\textbf{!!!~Giray:} #1}}
    
\fi

\newif\ifiscarevision
\iscarevisionfalse
\ifiscarevision 
    \newcommand{\revD}[1]{\textcolor{heliotrope}{#1}}
    \newcommand{\revE}[1]{\textcolor{rufous}{#1}}
    \newcommand{\revCommon}[1]{\textcolor{blue}{#1}}

    \newcommandx{\changeCM}[2][1=]{\todo[linecolor=blue,backgroundcolor=blue!25,bordercolor=blue,#1,size=\scriptsize]{\revCommon{\textbf{#2}}}}
    
    \newcommandx{\changeC}[2][1=]{\todo[linecolor=safetyorange,backgroundcolor=safetyorange!25,bordercolor=safetyorange,#1,size=\scriptsize]{\revC{\textbf{#2}}}}
    
    \newcommandx{\changeD}[2][1=]{\todo[linecolor=heliotrope,backgroundcolor=heliotrope!25,bordercolor=heliotrope,#1,size=\scriptsize]{\revD{\textbf{#2}}}}
    
    \newcommandx{\changeE}[2][1=]{\todo[linecolor=rufous,backgroundcolor=rufous!25,bordercolor=rufous,#1,size=\scriptsize]{\revE{\textbf{#2}}}}
\else
    \newcommand{\revD}[1]{\textcolor{black}{#1}}
    \newcommand{\revE}[1]{\textcolor{black}{#1}}
    \newcommand{\revCommon}[1]{\textcolor{black}{#1}}

    \newcommandx{\changeCM}[2][1=]{\todo[disable,#1]{#2}}
    \newcommandx{\changeA}[2][1=]{\todo[disable,#1]{#2}}
    \newcommandx{\changeB}[2][1=]{\todo[disable,#1]{#2}}
    \newcommandx{\changeC}[2][1=]{\todo[disable,#1]{#2}}
    \newcommandx{\changeD}[2][1=]{\todo[disable,#1]{#2}}
    \newcommandx{\changeE}[2][1=]{\todo[disable,#1]{#2}}
\fi

\newif\ifcut
\cutfalse

\ifcut
   \newcommand{\gfcut}[1]{} 
\else
    \newcommand{\gfcut}[1]{\textcolor{red}{\sout{#1}}}
\fi

\newif\ifiscasubmission
\iscasubmissiontrue
\ifiscasubmission
    
    \newcommand{\gfbisca}[1]{}

\else
    \newcommand{\gfbisca}[1]{\textcolor{blue}{\textit{GF: #1}}}

\fi

\newif\ifsubmission
\submissiontrue
\ifsubmission

    \newcommand{\jgl}[1]{}

    \newcommand{\gfb}[1]{}
    \newcommand{\mayank}[1]{}
    
    \newcommand{\agy}[1]{#1}
    \newcommand{\agycomment}[1]{}
\else
    \newcommand{\jgl}[1]{\textcolor{brickred}{\textit{JGL: #1}}}
    \newcommand{\gfb}[1]{\textcolor{blue}{\textit{GF: #1}}}
    \newcommand{\todo}[1]{\textcolor{red}{\textbf{TODO: #1}}}
    
    \newcommand{\mayank}[1]{\textcolor{green}{\textit{Mayank: #1}}}

    \newcommand{\agy}[1]{\textcolor{orange}{#1}}
    \newcommand{\agycomment}[1]{\agy{\textbf{[@gy:} #1\textbf{]}}}

\fi

\newcommandx{\unsure}[2][1=]{\todo[linecolor=red,backgroundcolor=red!25,bordercolor=red,#1, size=\tiny]{#2}}
\newcommandx{\change}[2][1=]{\todo[linecolor=blue,backgroundcolor=blue!25,bordercolor=blue,#1,size=\tiny]{\textbf{#2}}}
\newcommandx{\feedback}[2][1=]{\todo[linecolor=yellow,backgroundcolor=yellow!25,bordercolor=yellow,#1]{#2}}
\newcommandx{\improvement}[2][1=]{\todo[linecolor=Plum,backgroundcolor=Plum!25,bordercolor=Plum,#1]{#2}}
\newcommandx{\thiswillnotshow}[2][1=]{\todo[disable,#1]{#2}}
\newcommandx{\completedRevision}[2][1=]{\todo[disable,backgroundcolor=red,#1]{#2}}
\newcommandx{\dataSource}[2][1=]{\todo[disable,backgroundcolor=red,#1]{#2}}
\newcommandx{\info}[2][1=]{\todo[linecolor=ddgreen,backgroundcolor=ddgreen!25,bordercolor=ddgreen,#1, size=\tiny]{#2}}

\newcommand{\boxbegin} {
	\begin{tcolorbox}[enhanced, frame hidden, colback=gray!50, breakable]
}

\newcommand{\boxend} {
	\end{tcolorbox}
}

\definecolor{lightblue}{rgb}{0.980, 0.956, 0.623}
\sethlcolor{lightblue}

\newcommand{\yboxbegin} {
	\begin{tcolorbox}[breakable, enhanced, frame hidden,
	enlarge top by=-0.25cm,
   enlarge bottom by=-0.1cm,
	colback=yellow!50]
}

\newcommand{\yboxend} {
	\end{tcolorbox}
}

\makeatletter
\patchcmd{\SOUL@ulunderline}{\dimen@}{\SOUL@dimen}{}{}
\patchcmd{\SOUL@ulunderline}{\dimen@}{\SOUL@dimen}{}{}
\patchcmd{\SOUL@ulunderline}{\dimen@}{\SOUL@dimen}{}{}
\newdimen\SOUL@dimen
\makeatother

\usepackage{amsmath,amsfonts}
\usepackage{algorithmic}
\usepackage{tikz}
\usepackage{graphicx}
\usepackage{textcomp}
\usepackage{xcolor}
\usepackage{booktabs}
\usepackage{siunitx}
\usepackage{algorithm2e}
\usepackage{multirow}
\usepackage{fancyhdr}
\usepackage{float}
\usepackage{setspace}
\usepackage{listings}
\usepackage{balance}
\usepackage{wrapfig}
\usepackage{marginnote}
\usepackage{caption}
\usepackage{boldline} % for bold table borders
\usepackage{colortbl} % for table background color
\usepackage{floatflt}
\usepackage{tabularray}
\usepackage{rotating}
\usepackage{eso-pic}
\usepackage{noindentafter}
\usepackage{datetime}
\usepackage[]{hyphenat}
\usepackage{tikzducks}
\usepackage{makecell}
\usepackage[absolute,overlay]{textpos}
\usepackage{everypage}% to add the hook on every page
\usepackage{microtype}
\usepackage{algorithmic}
\usepackage{graphicx}
\usepackage{textcomp}
\usepackage{xcolor}
\usepackage{fancyhdr}
\usepackage{multirow}
\usepackage{bigdelim}
\usepackage{subcaption}
\usepackage{booktabs}
\usepackage{setspace}
\usepackage{lipsum} 
\usepackage[colorinlistoftodos,prependcaption,textsize=tiny,textwidth=12mm]{todonotes}

\ifcameraready
\else
    \fancypagestyle{firstpage}
    {
        \fancyhead{}
        \fancyhead[C]{\textcolor{red}{CONFIDENTIAL DRAFT -- DO NOT DISTRIBUTE -- TO APPEAR IN HPCA'24} \\ \textcolor{MidnightBlue}{\emph{Version \versionnum~---~\today, \ampmtime}}  \hrule }
    }
\fi

\makeatletter
\def\bstctlcite{\@ifnextchar[{\@bstctlcite}{\@bstctlcite[@auxout]}}
\def\@bstctlcite[#1]#2{\@bsphack
 \@for\@citeb:=#2\do{%
   \edef\@citeb{\expandafter\@firstofone\@citeb}%
   \if@filesw\immediate\write\csname #1\endcsname{\string\citation{\@citeb}}\fi}%
 \@esphack}
\makeatother

\usepackage{xspace}
\usepackage{tikz}

\definecolor{denim}{rgb}{0.08, 0.38, 0.74}
\definecolor{darkblue}{rgb}{0.0, 0.18, 0.39}
\definecolor{green}{rgb}{0.73, 0.83, 0.69}
\definecolor{teal}{rgb}{0.0, 0.502, 0.502}
\definecolor{mauve}{rgb}{0.58,0,0.82}

\definecolor{darkspringgreen}{rgb}{0.09, 0.45, 0.27}
\definecolor{tangerine}{rgb}{0.95, 0.52, 0.0}
\definecolor{pinkred}{rgb}{0.92, 0.05, 0.31}
\newcommand\proposal{MARS\xspace}

\newcommand\gp{GenPIP\xspace}

\newcommand\RSGA{RSGA\xspace}

\newcommand\rh{RawHash\xspace}
\newcommand\rhtwo{RawHash2\xspace}
\newcommand\isp{In-Storage-Processing\xspace}
\newcommand\ISP{ISP\xspace}

\newcommand\PIM{PIM\xspace}

\newcommand*\circled[1]{\tikz[baseline=(char.base)]{
            \node[shape=circle,draw,inner sep=1pt] (char) {#1};}}
\newcommand*\circledgray[1]{\tikz[baseline=(char.base)]{
            \node[shape=circle,draw,inner sep=1pt,fill=lightgray] (char) {#1};}}
\newcommand{\betterRH}[1]{\textcolor{black}{#1}}
\newcommand\rg{\revB{\texttt{MARS}}\xspace}
\newcommand\rgCPU{\texttt{MS-CPU}$_{Fixed}$\xspace}
\newcommand\rgCPUFP{\texttt{MS-CPU}$_{Float}$\xspace}
\newcommand\rhCPU{\texttt{RH2}\xspace}
\newcommand\bc{\texttt{BC}\xspace}
\newcommand\rgSIMDRAM{\texttt{MS-SIMDRAM}\xspace}
\newcommand\rgEXT{\texttt{MS-EXT}\xspace}
\newcommand\genpip{\texttt{GenPIP}\xspace}

\makeatletter
\newcommand\semiHuge{\@setfontsize\semiHuge{23.5}{28.2}}
\g@addto@macro{\normalsize}{%
  \setlength{\abovedisplayskip}{2pt plus 1pt minus 1pt}
  \setlength{\belowdisplayskip}{2pt plus 1pt minus 1pt}
  \setlength{\intextsep}{2pt plus 1pt minus 1pt}
  \setlength{\textfloatsep}{3pt plus 1pt minus 1pt}
  \setlength{\dbltextfloatsep}{3pt plus 1pt minus 1pt}
  \setlength{\skip\footins}{4pt plus 1pt minus 1pt}
}
\makeatother

\newcommand{\squishlist}{
 \begin{list}{$\circ$}
  { \setlength{\itemsep}{0pt}
     \setlength{\parsep}{0pt}
     \setlength{\topsep}{3pt}
     \setlength{\partopsep}{0pt}
     \setlength{\leftmargin}{1em}
     \setlength{\labelwidth}{1em}
     \setlength{\labelsep}{0.5em} } }

\newcommand{\squishend}{
  \end{list}  }

\definecolor{mint}{rgb}{0.24, 0.71, 0.54}

\newcommand\kon[1]{{\color{black}{#1}}}

\newcommand\rev[1]{{\color{black}{#1}}}

\newcommand\revA[1]{{\color{black}{#1}}}
\newcommand\revB[1]{{\color{black}{#1}}}
\newcommand\revC[1]{{\color{black}{#1}}}
\def\todos{0}

\begin{document}
\bstctlcite{IEEEexample:BSTcontrol}

\title{\proposal: Processing-In-\textit{M}emory \textit{A}cceleration of \\ \textit{R}aw Signal Genome Analysis Inside the \textit{S}torage Subsystem}

\author{
Melina Soysal$^\dagger$~\quad
Konstantina Koliogeorgi$^\dagger$~\quad
Can Firtina$^\dagger$~\quad
Nika Mansouri Ghiasi$^\dagger$\\
Rakesh Nadig$^\dagger$~\quad
Haiyu Mao$^\star$~\quad
Geraldo F. Oliveira$^\dagger$\\
Yu Liang$^\dagger$~\quad
Klea Zambaku$^\dagger$~\quad
Mohammad Sadrosadati$^\dagger$~\quad 
Onur Mutlu$^\dagger$\vspace{10pt}
\\
$^\dagger$~\emph{ETH Zürich} \qquad 
$^\star$~\emph{King's College London} \\
\vspace{5pt}
}

\ifcameraready
 \thispagestyle{plain}
\else
  \thispagestyle{firstpage}
\fi
\pagestyle{plain}

\maketitle

\glsresetall

\begin{abstract}
\noindent Conventional genome analysis relies on translating the noisy raw electrical signals 
generated by DNA sequencing technologies 
into nucleotide bases (i.e., A, C, G, and T) through a computationally-intensive process called basecalling.
Raw signal genome analysis (\RSGA) has emerged as a promising approach towards enabling real-time genome analysis by \revA{directly} analyzing raw electrical signals \revA{without} the need for basecalling.
However, rapid advancements in sequencing technologies make it increasingly difficult for software-based \RSGA to match the throughput of raw signal generation.
Hardware-based RSGA acceleration has the potential to bridge the gap between software-based \RSGA and sequencing throughput.

\rev{This paper demonstrates} that while (i) conventional hardware acceleration techniques (e.g., specialized ASICs) in tandem with (ii) memory-centric approaches (e.g., Processing-In-Memory) can significantly accelerate \RSGA, the high \rev{volume} of genomic data \rev{greatly} shifts the performance and energy bottleneck from computation to I/O data movement.
As sequencing throughput increases, I/O overhead becomes the dominant contributor \rev{to both} runtime and energy consumption, limiting the scalability of \revA{both} \rev{processor-centric and main-memory-centric} accelerators.
Therefore, there is a pressing need to design a high-performance, energy-efficient system for RSGA that can both alleviate the data movement bottleneck and provide large acceleration capabilities.

We propose \proposal, a storage-centric system that leverages the heterogeneous resources available within modern storage systems \rev{(e.g., storage-internal DRAM, storage controller, \revA{flash chips})} alongside their large storage capacity to tackle \rev{\emph{both}} data movement and computational overheads of \RSGA in an area-efficient and low-cost manner. 
\proposal accelerates \RSGA through a novel hardware/software co-design approach \revA{using} three \revA{major techniques}.
First, \proposal modifies the \RSGA pipeline \rev{via} a previously unexplored combination of two filtering mechanisms and a quantization scheme, reducing hardware demands and optimizing for in-storage execution.
Second, \proposal accelerates the modified \RSGA steps directly within the storage device by leveraging both Processing-Near-Memory and Processing-Using-Memory paradigms, tailored to the internal architecture of the storage system. 
Third, \proposal orchestrates the execution of all steps via a streamlined control and data flow to fully exploit in-storage parallelism and minimize data movement.
\revB{Our evaluation shows that \proposal outperforms basecalling-based software and hardware-accelerated  state-of-the-art read mapping pipelines by 93$\times$ and 40$\times$, on average across different datasets, while reducing their energy consumption by 427$\times$ and 72$\times$.
\proposal improves the performance of state-of-the-art \RSGA-based read mapping pipeline by 28$\times$ while reducing its energy consumption by 180$\times$ on average across different datasets.}

\end{abstract}

\section{Introduction}\label{sec:Intro}

Identifying and analyzing an organism's DNA sequence, \rev{i.e.,} \emph{genome analysis}, has led to important advances in areas such as personalized medicine~\cite{ashley_precision_2016,flores_p4medicine_2013, sweeney_rapid_2021}, outbreak tracing~\cite{Bloom_nextgen_2021,yelagandula_multiplexed_2021}, and evolutionary biology~\cite{romiguier_comparative_2014,ellegren_determinants_2016,Prado-Martinez_greatape_2013}. 
\textit{Genome sequencing} is the experimental process of determining the nucleotide sequence of an organism’s DNA. 
As current technologies cannot \revA{generate a single long sequence for an entire genome}, DNA is first fragmented into short sequences, called \emph{reads}, which serve as input to computational \revB{analyses}~\cite{alkan2009personalized,firtina_rawhash_2023,zhang_sigmap_2021,kovaka_uncalled_2021,Bao2021Squigglenet,bonito,li_minimap2_2018,Poplin2018universal,lindegger2023rawalign,Li_mapping_2008} to reconstruct the genome and extract biological insights.
\rev{The analysis typically starts with \emph{mapping} reads to a known \emph{reference genome}~\cite{alser_accelerating_2020,alser_technology_2020}, followed by identifying mutations and other genetic variations~\cite{firtina_rawhash_2023,zhang_sigmap_2021,kovaka_uncalled_2021,Bao2021Squigglenet,bonito,bingol2021gatekeeper,li_minimap2_2018,Poplin2018universal,lindegger2023rawalign,Li_mapping_2008} during downstream analyses.}

\rev{\revA{Nanopore sequencing technology}~\cite{jain_nanopore_2018,zhang_sigmap_2021,cali_nanopore_2017,cherf2012automated,laszlo2013detection,laszlo2014decoding,jain_minion_2016} enables DNA sequencing by passing DNA strands through nano-scale pores, known as \emph{nanopores}, and measuring the resulting fluctuations in electrical current.
%These current measurements, referred to as \emph{raw signals}, form the basis for downstream analysis.
These current fluctuations, referred to as \emph{raw signals}, correspond to distinct sequences of DNA nucleotides and form the basis for downstream analyses.}
The small dimensions of the nanopores enable sequencing in compact devices~\cite{jain_minion_2016}, paving the way for portable, scalable, and low-cost~\cite{shendure_dna_2017} sequencing for a wide range of applications, including outbreak tracing and disease diagnosis~\cite{Greninger2015rapid,Kafetzopoulou2018assesment}.
Nanopore sequencers' rapid adoption is further driven by their unique capability of early termination of sequencing when further data is no longer \rev{needed~\cite{loose_real-time_2016, payne_readfish_2021}, \textbf{reducing the sequencing time and cost} and enabling \textbf{real-time} analysis~\cite{firtina_rawhash_2023,shih_haru_2023}.}

\rev{Typical genome analysis pipelines first translate noisy raw electrical signals into sequences of nucleobase characters through a process called basecalling~\cite{bonito,huang_sacall_2020,singh_framework_2024,cali_nanopore_2017}. Subsequent downstream analyses are then performed on these text-based sequences.
However, basecalling is computationally intensive and represents a major bottleneck for real-time analysis, as it relies heavily on} \revA{sophisticated} \rev{deep learning models}~\cite{dunn_squigglefilter_2021,mao_genpip_2022,singh_framework_2024,zhang_sigmap_2021}.
Given the increasing demand for real-time processing, there is a pressing need for developing fundamentally new algorithmic approaches to keep up with the rapid advances in nanopore sequencing in terms of performance, energy consumption, and cost~\cite{Wang_nanopore_2021, dunn_squigglefilter_2021,firtina_rawhash_2023,firtina_rawhash2_2023,lindegger2023rawalign}.

\rev{\revB{\textbf{Raw signal genome analysis (\RSGA)}}  \cite{firtina_rawhash_2023,zhang_sigmap_2021,shih_haru_2023,firtina_rawhash2_2023,kovaka_uncalled_2021,lindegger2023rawalign,firtina_rawsamble_talk_2024,dunn_squigglefilter_2021,Bao2021Squigglenet,payne_readfish_2021,edwards_real-time_2019,sadasivan_rapid_2023,mikalsen_coriolis_2023,shivakumar_sigmoni_2023} has been proposed as a new paradigm that bypasses traditional basecalling by operating \revA{\emph{directly}} on raw electrical signals. Instead of translating signals into nucleotide sequences, \RSGA analyzes the raw signals themselves to perform genomic tasks such as read mapping and variant detection. 
\RSGA can complement basecalling by serving as a lightweight pre-basecalling filter~\cite{cavlak_targetcall_2024} to reduce redundant basecalling operations or even replace basecalling entirely by directly analyzing raw signals \textbf{in real-time}, without translating them to nucleotide sequences first\revA{~\cite{dunn_squigglefilter_2021,firtina_rawhash_2023,firtina_rawhash2_2023,lindegger2023rawalign}.}
\RSGA can lead to
%\todo{todo:check claim}on-par accuracy and 
\revA{more comprehensive}~\cite{workman_nanopore_2019,kovaka_uncalled_2021,lindegger2023rawalign} genome analysis as it preserves richer sequencing information in the raw signals~\cite{lindegger2023rawalign,Wick2019Performance,wan_beyond_2022,rand_mapping_2017,simpson_detecting_2017,stephenson_direct_2022,alexandra_sneddon_real-time_2022}.
These key benefits have fueled rapid research progress in the field of \RSGA~\cite{firtina_rawhash_2023,zhang_sigmap_2021,shih_haru_2023,firtina_rawhash2_2023,kovaka_uncalled_2021,lindegger2023rawalign,firtina_rawsamble_talk_2024,dunn_squigglefilter_2021,Bao2021Squigglenet,payne_readfish_2021,edwards_real-time_2019,sadasivan_rapid_2023,mikalsen_coriolis_2023,shivakumar_sigmoni_2023}, opening new directions such as direct alignment~\cite{lindegger2023rawalign,kovaka2024uncalled4,dunn_squigglefilter_2021} and \textit{de novo} assembly~\cite{firtina_rawsamble_talk_2024} on raw signals.}

As advancements in sequencing technologies continue at a rapid pace, 
scalability challenges arise, placing increasing pressure on software-based \RSGA to match the throughput of raw signal generation and meet the real-time requirements. 
To bridge the widening gap between sequencing throughput and downstream analysis, \rev{hardware acceleration} is required to either process larger data volumes with the same computational resources or reduce execution time and energy consumption.
\revB{Research efforts target the computational bottlenecks of \RSGA by using GPUs \revA{(e.g.,~\cite{Bao2021Squigglenet,sadasivan_accelerated_2023, icdm10,Gamaarachchi2020_f5c,guo2019hardware})} or co-designing algorithms with specialized hardware architectures and ASICs (e.g.,~\cite{dunn_squigglefilter_2021, shih_haru_2023, sundaresan1992vlsi,10015864,samarasinghe2021energy})}.
While these approaches effectively reduce computational overhead, they largely overlook the impact of \rev{I/O data movement from the storage subsystem} on the \emph{end-to-end} \RSGA pipeline.
Our motivational analysis \rev{(\S\ref{sec:motivation})} shows that as \rev{the computational bottlenecks of \RSGA} are accelerated, the contribution of I/O becomes dominant and ultimately emerges as the primary bottleneck in the end-to-end analysis.
For instance, as the accelerator speedups increase, the adverse impact of the storage subsystem dominates the accelerated end-to-end execution latency, reaching up to 78\% of total execution time for large genomes (see \rev{\S\ref{sec:motivation-why-isp}}). 
This motivational study highlights the need for an architecture for \RSGA that (i)~alleviates \revA{the large} data movement overhead, (ii)~accelerates the computational \revA{steps} of \RSGA, and (iii)~scales to the large volumes of genomic datasets.

\textbf{Our goal} in this work is to design a high-performance, energy-efficient, and scalable system for \RSGA by effectively addressing \emph{both} the data movement and computation overheads of the end-to-end \RSGA pipeline for read mapping.
%across a wide range of genomic datasets. 
Our key idea is to design a \rev{\emph{storage-centric} system that leverages the \emph{heterogeneous compute-capable resources} (e.g., SSD internal DRAM, SSD controller)}, alongside the large storage capacity available within modern storage systems to alleviate I/O data movement and \rev{computational bottlenecks within the \RSGA pipeline} in an area-efficient and low-cost manner. 
To this end, we propose \emph{\proposal} (Processing-In-\textbf{M}emory \textbf{A}cceleration of \textbf{R}aw Signal Genome Analysis Inside the \textbf{S}torage Subsystem), the first \isp (ISP) design combining Processing-Using-DRAM and Processing-Near-DRAM \emph{\textbf{within}} a storage system.

\textbf{Challenges.} 
%\todo{I think here we should mention the challenges of isp-pim codesign}
Despite ISP's promising potential, designing a storage-centric system for \RSGA presents several key challenges.
First, \RSGA steps (e.g., event detection, seeding and chaining) exhibit high memory demands and irregular data access patterns. In contrast, SSDs lack architectural support for fine-grained (i.e., small-size) memory operations and are optimized for sequential access to fully utilize the high flash memory channel bandwidth inside the SSD.
Second, exploiting heterogeneous resources and computation capabilities within the \revB{storage system} introduces a complex design space and a rich set of tuning parameters.
Third, deploying the end-to-end \RSGA pipeline consisting of multiple steps inside the storage \revC{system} creates contention over shared resources, requiring careful coordination and isolation.
Addressing these challenges necessitates a carefully-constructed design to ensure a synergistic and efficient orchestration of the available in-storage resources.

We address these challenges through a novel hardware/software co-design approach that modifies and enables \RSGA computational primitives to leverage in-storage execution capabilities, while carefully taking into account storage system constraints.
First, we propose two software modifications: 
(1)~a novel combination of two filtering mechanisms~\cite{firtina_rawhash2_2023,li_minimap2_2018,liu_fast_2016,liao_subread_2013} that selectively remove redundant or low-quality candidate matches between the input and reference genomes early in the \RSGA pipeline, reducing both computational workload and intermediate data storage requirements and 
(2)~\revB{an arithmetic conversion} scheme that reduces the precision of intermediate signal representations to lower storage and computation overheads, carefully placed in the \RSGA pipeline to preserve accuracy.
\revB{Second, we augment the storage system's functionality to support the \RSGA pipeline by placing accelerators for individual steps in different parts of the storage subsystem, leveraging different ~\emph{`Processing-In-Memory'} paradigms:} 
\revB{(i)~~inside the memory array of the storage-internal DRAM through the \emph{`Processing-Using-DRAM'} approach, 
(ii)~~near the subarrays of the storage-internal DRAM using 
 the \emph{`Processing-Near-DRAM'} approach and
 (iii)~~inside the storage controller via the \emph{`Processing-Near-DRAM'} approach, which operates on data fetched from the storage-internal DRAM.  
\proposal orchestrates these individual components through a unified control and data flow that minimizes data movement and \revB{efficiently} exploits the available bandwidth between them.}

We evaluate \proposal-based read mapping in terms of accuracy, latency and energy consumption \rev{across five diverse genomic input datasets from different species.}
We compare our design against four state-of-the-art software and hardware baselines using both \RSGA and basecalling-based approaches and make \revB{four} major observations.
First, \proposal outperforms the state-of-the-art CPU-based \RSGA implementation for read mapping ~\cite{firtina_rawhash2_2023} by 28$\times$, on average, across all datasets while improving the energy consumption by 180$\times$ on average. 
Second, \proposal provides an average speedup of 93$\times$ over a hybrid CPU/GPU-accelerated basecalling-based pipeline~\cite{dorado,li_minimap2_2018}, while improving energy consumption by 427$\times$ on average. 
%Third, \proposal achieves a significant speedup of 1.62$\times$ compared to state-of-the-art \RSGA hardware accelerator \sqf~\cite{dunn_squigglefilter_2021}, that only supports analysis of small viral genomes.
Third, \proposal is superior to \gp~\cite{mao_genpip_2022}, a state-of-the-art Processing-In-Memory-based read mapping system relying on basecalling, achieving a speedup of 40$\times$ and energy savings of 72$\times$ on average across all five datasets. 
Fourth, \proposal provides analysis accuracy \revA{\emph{on par}} with the conventional basecalling-based pipeline. 
%Fifth, \proposal's performance and energy benefits come at a low area cost, as \proposal introduces less than 1.7\% hardware area overhead in a modern SSD device\todo{We checked. Please see evaluation.}.

\noindent This work makes the following \textbf{key contributions}:
%\vspace{-.2em}
\squishlist
    \item It is the first work to \revC{demonstrate the I/O} bottleneck of hardware-accelerated Raw Signal Genome Analysis (\RSGA) and propose \isp of \RSGA.
    \item We propose \proposal, the \textit{first} \isp system for \RSGA, which mitigates both I/O data movement and computational overheads through a tightly integrated hardware/software co-design.  
    \item To our knowledge, \proposal is the first architecture to integrate \revC{\emph{multiple Processing-In-Memory paradigms}} within the storage \revC{system}. We implement accelerators \emph{inside} the SSD's DRAM, \emph{near the subarrays} of the SSD's DRAM as well as \emph{inside the SSD controller} leveraging both Processing-Using-DRAM and Processing-Near-DRAM paradigms to efficiently enable diverse \RSGA computation primitives.
    \item We extensively compare \proposal to state-of-the-art software- and hardware baselines that use both \RSGA and basecalling. We show that \proposal improves performance over software and hardware-accelerated state-of-the-art \ifnum\todos=1\todo{The reason why we use read mapping is to not invoke questions about comparing quantitatively with alignment tools eg squigglefilter.}\fi read mapping pipelines by a factor of 93$\times$ and 40$\times$ while reducing their energy consumption by 427$\times$ and 72$\times$ on average across five real-world datasets.
\squishend

%The reason why we use read mapping pipeline here is that we also compare against pipelines that are not RSGA-based (GenPIP). We ensured however, that we are as precise if possible --> We always use RSGA pipeline and only use read mapping pipeline if we include non-RSGA based pipelines
\section{Background}
\label{sec:background}

% We provide a concise overview of \rsga, NAND flash-based SSD, and Data-Centric Computing that is essential for understanding the subsequent sections of the paper.

\subsection{Genome Analysis}
\label{sec:BG-RSGA}
\textbf{Raw Signal Genome Analysis.}
Nanopore sequencing can sequence \revC{relatively} long fragments of DNA~\cite{jain_nanopore_2018,zhang_sigmap_2021,cali_nanopore_2017,cherf2012automated,laszlo2013detection,laszlo2014decoding,jain_minion_2016}, called \emph{reads}, by measuring the electrical current changes caused when a DNA fragment traverses a tiny pore, called \emph{nanopore}.
%Longer fragments improve the quality of the analysis~\cite{jain_nanopore_2018}. 
The generated sequence data~\cite{jain_nanopore_2018,zhang_sigmap_2021,cali_nanopore_2017,cherf2012automated,laszlo2013detection,laszlo2014decoding,jain_minion_2016}, referred to as \emph{raw signals}, are then used in downstream analysis, e.g., for read mapping~\cite{firtina_rawhash_2023,kovaka_uncalled_2021,zhang_sigmap_2021,firtina_rawhash2_2023,Bao2021Squigglenet,shih_haru_2023} and alignment~\cite{lindegger2023rawalign,dunn_squigglefilter_2021,kovaka2024uncalled4} purposes.
In the conventional genome analysis approach, raw signals are first translated into sequences of nucleobase characters (i.e., A, C, G, T) during the basecalling process~\cite{alser_technology_2020,alser2022molecules,dorado,bonito,cavlak_targetcall_2024}, and then mapped to a reference genome to \rev{find similarities and differences~\cite{bonito,huang_sacall_2020,singh_framework_2024,xu2021fastbonito,zeng2020causalcall,singh2024rubicon}.
In contrast,} \RSGA eliminates the need for basecalling by directly operating on \revB{raw signals}~\cite{firtina_rawhash_2023,zhang_sigmap_2021,shih_haru_2023,firtina_rawhash2_2023,kovaka_uncalled_2021,lindegger2023rawalign,firtina_rawsamble_talk_2024,dunn_squigglefilter_2021,Bao2021Squigglenet,payne_readfish_2021,edwards_real-time_2019,sadasivan_rapid_2023}.

%\begin{figure}[tbh]
 % \centering
 % \includegraphics[width=0.9\columnwidth]{figures/Pipelines-Comparisons.pdf}
 % \caption{Comparison of the conventional genome analysis and the raw signal genome analysis pipeline.}
  %\label{fig:pipeline-comparison}
%\end{figure}

\begin{figure}[h]
  \centering
  \includegraphics[width=\columnwidth]{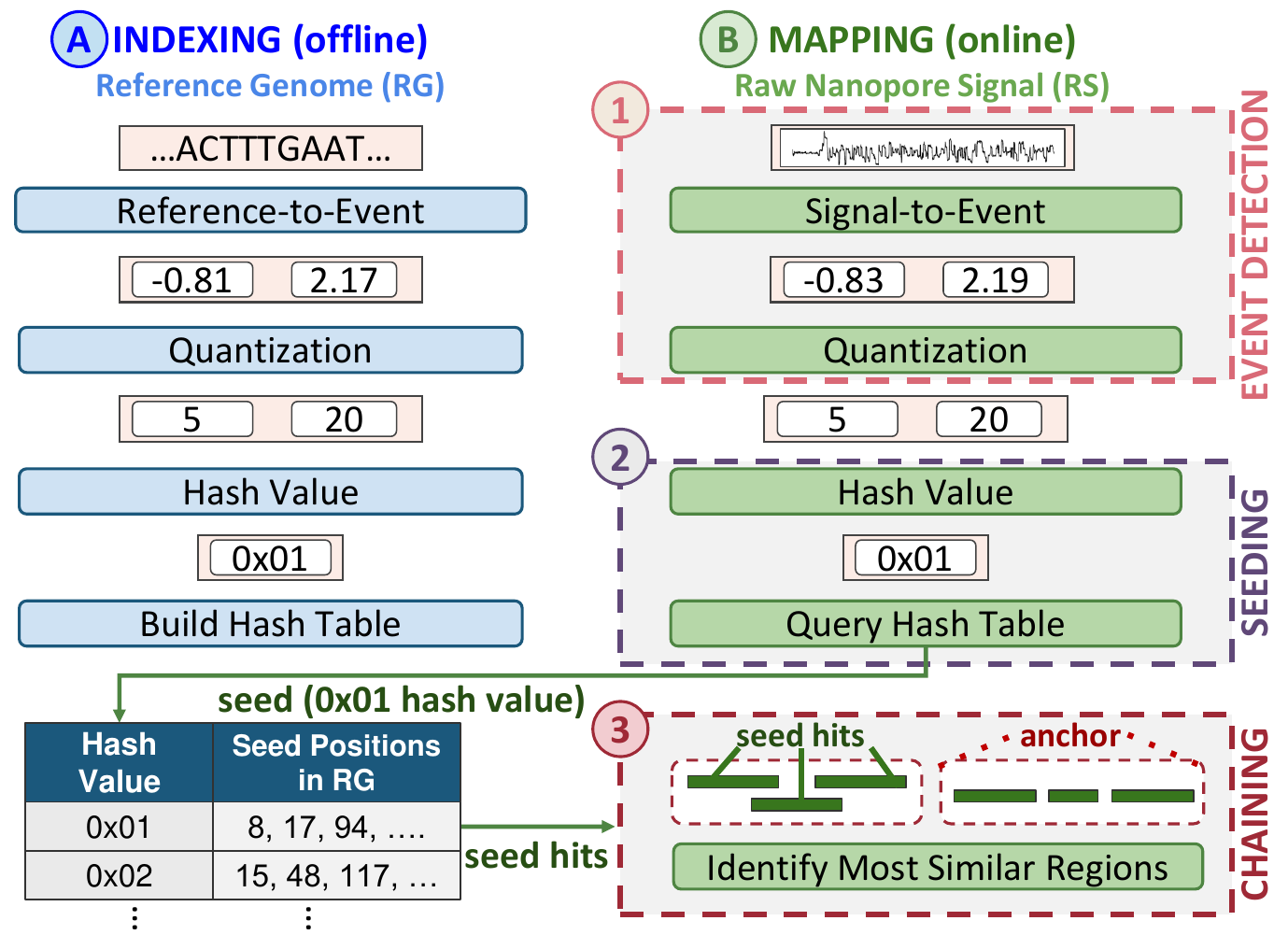}
  \caption{\revB{Overview of a typical \RSGA read mapping workflow based on a hash-table for indexing.}} 
  %denotes the position of the reference genome ($P_{RG}$) and the raw signal ($P_{RS}$).
  \label{fig:rawhash-flow}
\end{figure}

\RSGA requires comparing sequences from the reference genome \rev{with sequences derived from \revB{each input query}, i.e., the raw electrical signals generated by the sequencer for a given DNA sample.}
To enable this comparison, both reference subsequences and raw signals are converted into \emph{events}, i.e., a series of values corresponding to genomic subsequences of certain length.
These event sequences are then passed through a quantization step that accounts for sequencing noise and enables robust signal-domain comparisons between reference and input query.
A typical state-of-the-art \RSGA pipeline \rev{for read mapping}, illustrated in Fig.~\ref{fig:rawhash-flow}, consists of two main \rev{stages} \revC{~\circled{A}~Indexing} and \revC{~\circled{B}~Mapping}.

\vspace{3mm}

\vspace{3mm}

\noindent\rev{~\circled{A}~Indexing} (offline): The reference genome is converted into events through reference-to-event conversion and quantization. 
These events are then stored in an efficient data structure, e.g., a hash table, to enable fast lookup of matching signal patterns. 
\rev{~\circled{B}~Mapping} (online): This \rev{stage} maps raw signals to the reference genome using the previously constructed index.
\revA{The first step~\circled{1} in mapping} is \textit{event detection}, which performs the signal-to-event conversion of raw signals and applies quantization.
\revB{In the second step~\circled{2}}, called \textit{seeding}, consecutive events are grouped to generate hash values which represent signal segments known as \emph{seeds} and are used to query the reference index. 
Matching entries, referred to as \emph{seed hits}, represent candidate matches between the input and reference.
\revB{During the last step~\circled{3}}, called \textit{chaining}, seeds are sorted based on their positions in the reference genome. 
Seeds that are both spatially close and colinear, i.e., \revB{those that} maintain consistent relative positions in the reference and input query, are grouped into \textit{anchors}, which form the basis for constructing \textit{chains} representing high-confidence matching regions between the query and \revC{the} reference genome.

\begin{figure}[h]
  \centering
  \includegraphics[width=\columnwidth]{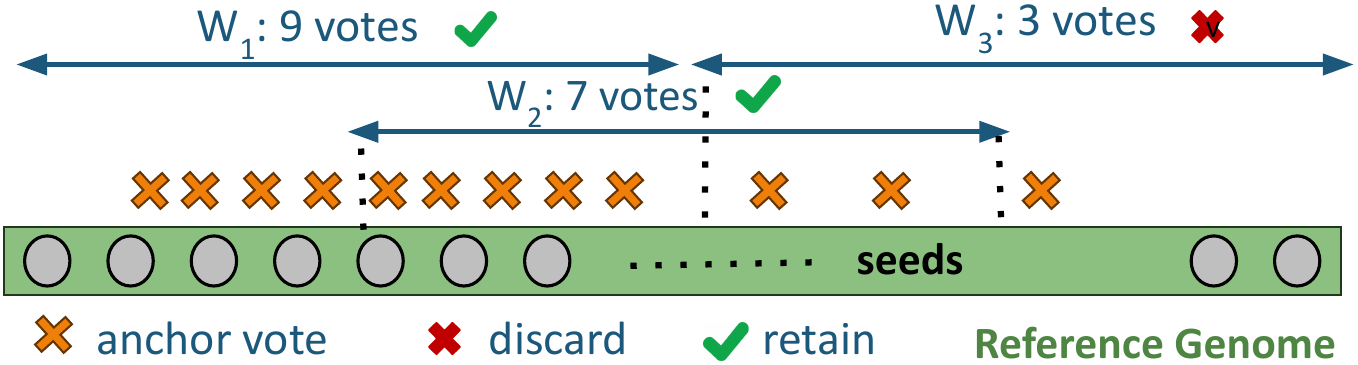}
  \caption{\revB{Overview of the seed-and-vote filtering technique for a threshold value of 5}.}
  \label{fig:seed-vote}
\end{figure}

\textbf{Filtering Techniques.}
\revB{Filtering techniques~\cite{alser2019shouji,alser2020sneakysnake,bingol2021gatekeeper,xin2013fasthash,alser2017gatekeeper,li_minimap2_2018,kim2018grim,rizk2010gassst,hach2014mrsfast,liao_subread_2013,laguna2020seed,xin2015shifted,xin2016optimal,myers1995chaining,alser2017magnet}} are extensively used in genome analysis pipelines to reduce the need for costly alignment operations by eliminating unlikely candidate matches early during the read mapping process.
One popular filtering approach, adopted in both conventional and \RSGA approaches, is \textit{frequency filtering}~\cite{li_minimap2_2018,firtina_rawhash2_2023}. \revB{The goal of frequency filtering is to identify and eliminate the seeds that cause a large number of seed hits in the reference genome. These frequent seed hits usually appear due to repetitions in the genome or hash collisions, which can cause ambiguity~\cite{firtina_genomic_2016} in read mapping and increase the computational cost of the subsequent steps~\cite{alser_technology_2020}, such as chaining. 
To eliminate these issues, these seed hits are \revC{\emph{not}} considered in the subsequent chaining stage, effectively reducing the computational load.}
A dataset-specific value defines the threshold for filtering out such frequent matches.
Another promising method is the \textit{seed-and-vote} filtering technique~\cite{liu_fast_2016,liao_subread_2013} that has been applied in conventional basecalling-based pipelines to discard anchors that are unlikely to generate valid alignments.
As shown in Fig.~\ref{fig:seed-vote}, the reference genome is partitioned into overlapping, equal-length windows $W_i$. Each anchor
%, representing a seed-level match between the reference genome and input query, 
votes for the window(s) it appears in (see orange $X's$ \revC{in Fig.~\ref{fig:seed-vote}}).
%A high number of votes per window indicates a region that contains correct alignments.
\revC{We define as \textit{voting threshold} the minimum number of votes per window, so that it contains correct alignments.  
A region whose vote count falls below this predefined threshold is excluded from further analysis, reducing the computational load of chaining.}
\vspace{1mm}
\vspace{1mm}

\subsection{SSD Architecture}

\begin{figure}[h]
  \centering
  \includegraphics[width=0.9\columnwidth]{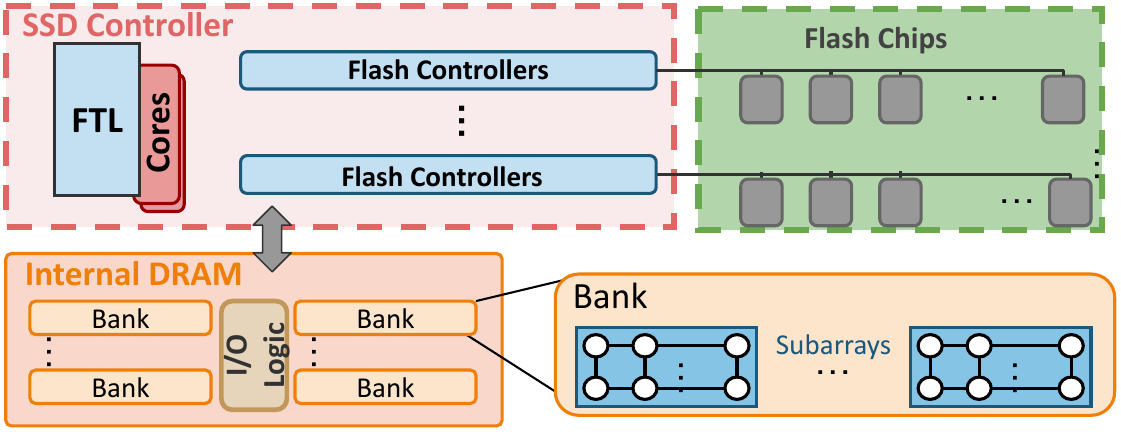}
  \caption{Organizational overview of a modern SSD.}
  \label{fig:ssd-arch}
\end{figure}

%NAND flash-based Solid State Drives (SSDs) are the prevalent storage medium due to their high performance, low power consumption, and shock resistance\rev{~\cite{dirik_isca_2009, mielke2017reliability, kryder2009after,do2013query,cho2013active,kim2018autossd,eshghi2018ssd,micheloni2010inside,cai2013threshold,cai2015data,cai2017error,park2022flash,kim2020evanesco,tavakkol2018flin}}.
Fig.~\ref{fig:ssd-arch} depicts the architecture of a typical modern NAND flash-memory-based Solid State Drive (SSD)~\cite{micheloni2013inside}, which consists of three main components: (1)~an array of NAND flash chips, 
%connected via flash channels to 
(2)~SSD controller, and (3)~DRAM.
%in SSD for caching metadata and frequently-requested data~\cite{gupta_dftl_2009}. We provide more details about components related to our work.
\vspace{3mm}
\vspace{3mm}

\noindent\textbf{NAND Flash Memory.}
NAND flash memory consists of multiple flash chips~\cite{cai2017error,agrawal2008design}, which are connected to the SSD controller via multiple parallel flash channels. 
Each flash chip typically contains one or more independent dies. Each die has multiple (e.g., 2 or 4) planes and each plane
contains thousands of blocks. A block includes hundreds to thousands of pages, each of which is 4–16 KiB in size.

%The internal bandwidth (BW) between NAND chips and the SSD controller referred to as flash channel BW, typically exceeds the external BW available between the SSD and the host system, called PCIe lane BW.  
%Recent enterprise SSD controllers~\cite{anandcontroller_2020} support 6.55GB/s external and 19.2GB/s internal BW, distributed over 16 channels operating at 1.2 GB/s each. 
%To bridge the performance gap between main memory and storage systems, modern SSDs adopt high-speed I/O interfaces.
%For instance, cutting-edge PCIe-Gen4 SSDs can achieve sequential read bandwidths of up to 8 GB/s (e.g., 7 GB/s in Samsung PM1735~\cite{samsungPM1735_2020}).

\noindent\textbf{SSD Controller.}
The SSD controller~\cite{cai2017error,cai2017errors,micheloni2013inside} consists of two primary components: (1)~multiple general-purpose cores running the SSD firmware, i.e., the \emph{flash translation layer (FTL)}, and (2)~per-channel hardware \emph{flash controllers}.
The FTL manages communication with the host system, maintains logical-to-physical (L2P) address mappings for read operations, handles internal I/O scheduling, and performs various SSD management functions to hide the complexities of NAND flash memory from the host processor.
Flash controllers handle (i) requests between the SSD controller and the flash chips and (ii)~error-correcting codes (ECC) for the NAND flash chips~\cite{cai2017error,cai2017errors,zhao2013ldpc,tanakamaru2013error}. 

\begin{figure}[h]
  \centering
  \includegraphics[width=\columnwidth]{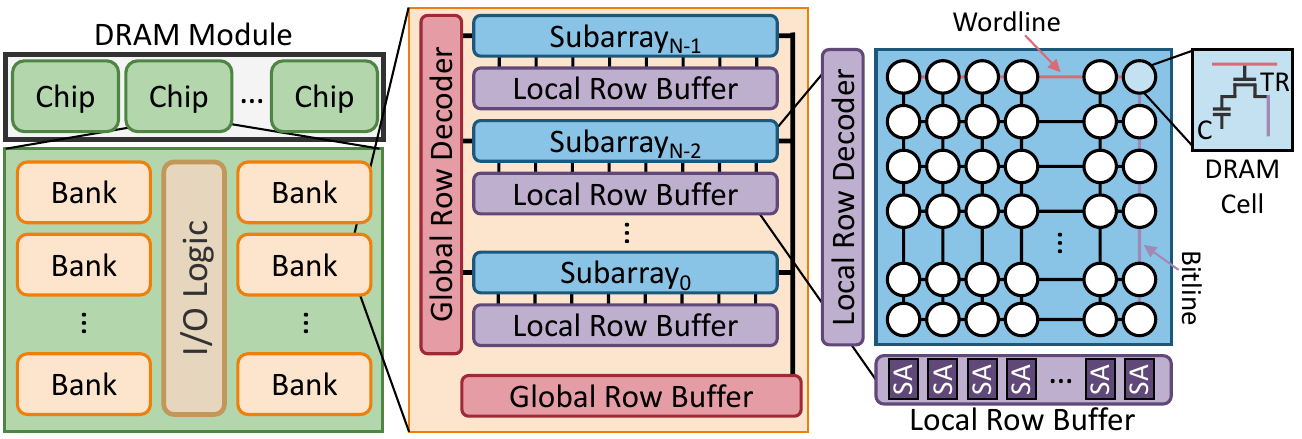}
  \caption{Organizational overview of a DRAM module.}
  \label{fig:dram-arch}
\end{figure}

\noindent\textbf{SSD-Internal DRAM}.
Modern SSDs employ DRAM to store metadata crucial for SSD management (e.g., L2P page mapping table) and to cache frequently accessed pages~\cite{gupta_dftl_2009,lim2010faster,zhou2015efficient,tavakkol2018mqsim,shin2009ftl,justinmezasigmetrics15}. 
Typically, the DRAM takes up 0.1\% of the SSD's capacity (e.g., 4GB LPDDR4 DRAM~\cite{ddr4} for a 4TB SSD~\cite{samsung860pro_2018}). 
As shown in Fig.~\ref{fig:dram-arch}, DRAM is organized in a hierarchical structure. At the highest level, a DRAM module comprises multiple chips, each containing several banks (e.g., 8-16), subdivided into multiple subarrays (e.g., 64-128).
A subarray is a 2D array of cells organized into multiple rows (e.g., 512-1024) and columns (e.g., 2-8 KB)~\cite{kim2018solar,lee2017design}. 
Cells in a row share a wordline while cells in the same
column share a bitline. The bitline is used to read from and write to the cells via the row buffer, which contains sense amplifiers (SA in Fig.\ref{fig:dram-arch}).

%To read data, we set the row address to copy the relevant row content into the local row buffer, \revB{i.e., sense amplifiers (SA in Fig.\ref{fig:dram-arch})}. The chosen column is then routed to the local data line via column selection lines.
\vspace{3mm}
\vspace{3mm}

\noindent\textbf{SSD I/O Bandwidth.} SSDs are characterized by the external and internal bandwidth (BW). 
\revC{The external BW, e.g., PCIe~\cite{PCIE4,eshghi2018ssd} lane BW, refers to the data transfer rate between the SSD and the host system and is determined by the number of PCIe lanes. 
In contrast, the internal BW refers to the bandwidth between the NAND flash chips and the SSD controller.}
%The external BW, e.g. PCIe lane BW, is the BW available between the SSD and the host system, whereas the internal BW is the BW between NAND flash chips and the SSD controller. 
The internal BW typically exceeds the external BW.
For example, recent enterprise SSD controllers~\cite{anandcontroller_2020} support 6.55GB/s external and 19.2GB/s internal BW, distributed over 16 channels operating at 1.2 GB/s each~\cite{kang2019vnand}. 
To bridge the performance gap between main memory and storage systems, modern SSDs integrate cutting-edge PCIe-Gen4 interfaces, e.g., 7 GB/s PCIe in Samsung PM1735~\cite{samsungPM1735_2020}.
%For instance, cutting-edge PCIe-Gen4 SSDs can achieve sequential read bandwidths of up to 8 GB/s (e.g., 7 GB/s in Samsung PM1735~\cite{samsungPM1735_2020}).

\section{Motivation}
\label{sec:motivation}

\subsection{Computational Requirements of \RSGA}
\label{sec:motivation-why-rsga}

\RSGA is a promising approach for bridging the performance gap between sequencing technologies, such as Nanopore sequencing~\cite{jain_nanopore_2018,zhang_sigmap_2021,cali_nanopore_2017,cherf2012automated,laszlo2013detection,laszlo2014decoding,jain_minion_2016}, and analysis times.
It can reduce the basecalling workload by serving as a pre-basecalling filtering approach ~\cite{cavlak_targetcall_2024} or enable real-time analysis by completely bypassing the costly deep-learning based basecalling step~\cite{dorado,bonito,huang_sacall_2020,singh_framework_2024,xu2021fastbonito,zeng2020causalcall,singh2024rubicon}. 
However, given the rapid growth of sequencing throughput, it becomes exceedingly challenging for software-based \RSGA to meet the requirements of real-time analysis~\cite{dunn_squigglefilter_2021}. 
The increasing number of flow cells and nanopores per flow cell lead to scalability challenges in processing generated data simultaneously and in real-time.
Real-time \RSGA~\cite{dunn_squigglefilter_2021,loose_real-time_2016,firtina_rawhash_2023,firtina_rawhash2_2023,Bao2021Squigglenet,zhang_sigmap_2021,shih_haru_2023}, particularly for large genomes and extensive data sets, requires medium to large-sized server-grade systems to meet the significant computational and memory needs~\cite{firtina_rawhash_2023,zhang_sigmap_2021,kovaka_uncalled_2021}.  
For example, mapping a human genome with \RSGA on our server-grade system (configuration in \S\ref{sec:Methodology}) requires 52 CPU threads and 128 GB DRAM capacity to meet the real-time analysis requirements of a single portable palm-sized sequencing device~\cite{firtina_rawhash2_2023}.
Recent state-of-the-art works~\cite{dunn_squigglefilter_2021, shih_haru_2023} meet real-time requirements for small genomes, but fail to scale to larger inputs due to the computationally costly operations of full-genome \emph{alignment}, which slows down the system at quadratic rates as the genome size increases.
To further understand the acceleration obstacles of \RSGA workflows and exploit the full potential for acceleration, a systematic analysis is required.

\begin{figure}[h]
  \centering
  \includegraphics[width=\columnwidth]{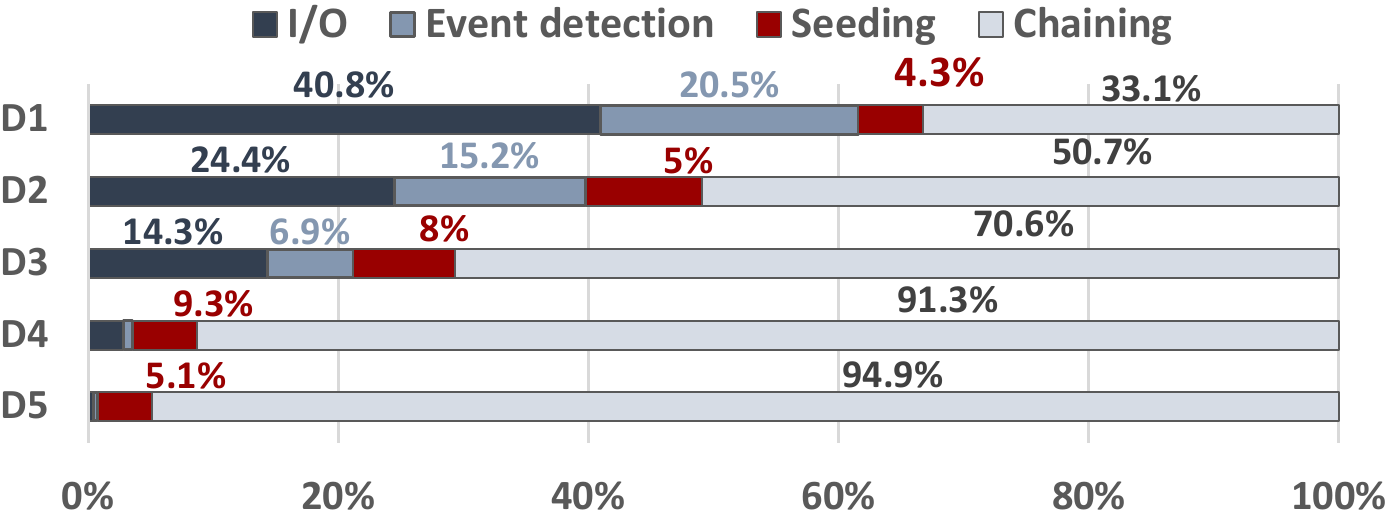}
  %{figures/Breakdown-Motivation.pdf}
  \caption{\rhtwo runtime breakdown for real-world genomic datasets, from smallest (D1) to largest (D5).}
  \label{fig:motivation-breakdown}
\end{figure}

We focus on \rhtwo~\cite{firtina_rawhash_2023}, the state-of-the-art \RSGA pipeline for read mapping that uses efficient quantization and a lightweight hash-based similarity search to scale to larger genomes. 
We choose \rhtwo as it introduces a highly efficient seed search mechanism, that leads to a better accuracy-throughput trade-off in comparison to prior \RSGA read mapping mechanisms, Sigmap~\cite{zhang_sigmap_2021}, UNCALLED~\cite{kovaka_uncalled_2021} and \rh~\cite{firtina_rawhash_2023}.
We execute \rhtwo on a high-end, latency-optimized SSD~\cite{samsungPM1735_2020} with a PCIe Gen4 interface (\texttt{PCIe})~\cite{PCIE4}. 
Fig.~\ref{fig:motivation-breakdown} shows the breakdown of \rhtwo into the steps described in \S\ref{sec:background} (i.e., event detection, seeding, chaining) as well as I/O overhead. 
We measure I/O overhead by executing the pipeline once with data fully preloaded in memory \revC{(i.e., without I/O overhead)}, and once \revC{with no data preloaded into memory (i.e., with full I/O overhead from storage).}
The difference in total runtime between the two runs reflects the I/O data movement time from \revC{SSD} to memory.
We \revC{use} five different datasets \revC{as inputs}, enumerated from the smallest (D1, viral SARS-CoV-2 genome) to the largest one (D5, human genome). 
For \emph{all} genome sizes, chaining is consistently a primary computational overhead, contributing between 33.1\% (D1) and 94.9\% (D5) of the total execution time.
Seeding takes up 4.3\%-9.3\% of the execution time. Event detection and I/O data overhead are considerable bottlenecks \revC{especially} for small datasets \revC{(D1, D2, D3), taking up to 20.48\% and 40.84\% of the execution time respectively}.
\vspace{2mm}
\vspace{2mm}

While no prior work has accelerated the full \RSGA read mapping pipeline end-to-end, \revC{several of its individual compute primitives have been the focus of hardware acceleration efforts~\cite{alser_accelerating_2020,Mutlu2023Accelerating,alser2022molecules,firtina2025enabling}.
\textbf{Chaining acceleration} has received significant attention in the literature. 
Researchers have employed GPUs~\cite{sadasivan2023accelerating,guo2019hardware} as well as FPGAs and custom hardware architectures~\cite{guo2019hardware,shih_haru_2023,10015864,liyanage2023efficient,gu2023gendp}, achieving performance improvements ranging from 5.4$\times$ to 277$\times$ compared to their respective software baselines.
More recently, novel computing paradigms have been explored to accelerate chaining, including \PIM architectures~\cite{chen2020parc} and RISC-V custom instructions~\cite{liyanage2024accelerating}.
\textbf{Seeding acceleration} has similarly been investigated. Custom hardware designs~\cite{10609692} and GPU-based implementations~\cite{seedgpu1,seedgpu2} have demonstrated the potential for significant performance gains. 
In particular, hash-based seeding has emerged as a promising target for in-memory acceleration, with several works proposing PIM-based solutions~\cite{huang2023casa,huangfu2019medal,zhang2021pimquantifier,jahshan2024majork,huangfu2020nest,zokaee2019finder}. 
For example, ~\cite{mao_genpip_2022} implements a ReRAM-based accelerator for hash-based seeding within basecalling pipelines, leveraging similar compute primitives as the seeding step in \RSGA.}
pLUTo \cite{Ferreira2022pLUTo}, an in-DRAM accelerator optimized for lookup-table (LUT) operations, is a promising approach for accelerating hash-based seeding and achieves speedups of up to 700$\times$ over CPU baselines for seeding-relevant workloads.

\subsection{\revB{Impact of Data Movement on Hardware Accelerated \RSGA}}
\label{sec:motivation-why-isp}
Despite the promising results of \revB{these standalone accelerators}, there are no mature end-to-end accelerated systems for \RSGA. 
Current works overlook the impact of storage I/O on the end-to-end accelerated system. 
As more \RSGA pipeline steps are accelerated to meet the real-time requirements and the growing throughput of modern sequencing devices, %time distribution across \RSGA read mapping will change drastically. 
the distribution of latency across \RSGA read mapping steps will change drastically.
We expect I/O to emerge as the dominant bottleneck in the end-to-end analysis as the computational steps are increasingly accelerated and thus minimized.

\begin{figure}[h]
\centering
\includegraphics[width=\columnwidth]{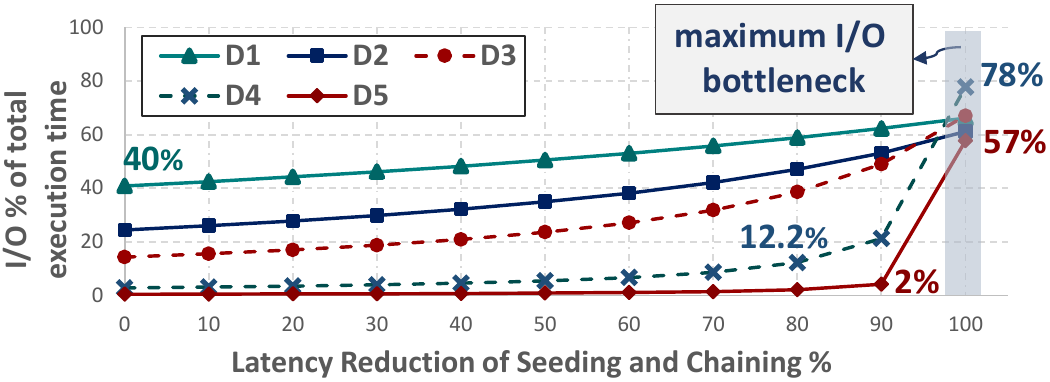}
\caption{Impact of I/O on overall execution time under increasing acceleration of \revC{computation} bottlenecks.}
\label{fig:motiv-model}
\end{figure}

We validate this hypothesis through a motivational experiment that analyzes \rhtwo~\cite{firtina_rawhash_2023} using the same real-world datasets and hardware setup as introduced in our previous experiment \revB{(\S \ref{sec:motivation-why-rsga})}.
\revB{We assume a scenario} that applies state-of-the-art accelerators to the two most frequently accelerated steps: seeding and chaining.
\revB{We model the latency of the accelerated workflow by incrementally reducing the latency of the seeding and chaining steps by 10\% until we reach 100\% total latency reduction, i.e., \revC{zero} execution time.} \revC{The results are shown in Fig.~\ref{fig:motiv-model}}\ifnum\todos=1\todo{named the grey area as the area with max io overhead to avoid characterizing the accelerators}\fi. 
\vspace{3mm}
\vspace{3mm}

Fig.~\ref{fig:motiv-model} shows how I/O data movement overhead \revC{progressively dominates end-to-end execution time as latency reduction increases}. We make the key observation that as latency reduction increases, the I/O \revB{data} overhead becomes the limiting factor across all datasets. 
In \revC{particular}, for small genome datasets (D1-D3), I/O overhead reaches up to 66\% of the total execution time. 
For larger genomes, I/O overhead remains modest until \revC{a} large latency \revC{reduction} of 90\%.
However, as \revC{computation} bottlenecks are further minimized, storage I/O emerges as the primary performance limiter.
\textbf{For example, I/O overhead accounts for 57\% and 78\% of the total execution time for D5 and D4 respectively when \revC{execution time of seeding and chaining is reduced by 100\%}.}
These results indicate that accelerating the seeding and chaining alone is insufficient, and that I/O \revB{data} movement \revB{from SSDs} becomes the dominant overhead in \revB{accelerated} \RSGA.

\subsection{Our Goal}

Based on our motivational analysis, acceleration of \RSGA is critical for achieving real-time genome analysis. 
While computational complexity is a key challenge, I/O data movement from SSDs becomes the dominant bottleneck across all genome datasets once computational steps are \revC{heavily} accelerated.
\textit{In-Storage processing} (\ISP) can, therefore, be a key enabler for designing a real-time system for \RSGA. 
Specifically, \ISP can uniquely address the I/O data movement bottleneck, manage the high volume of genomic data, and provide fine-grained parallelism to accelerate the computational steps.
However, designing an \ISP system for \RSGA is challenging due to architectural constraints of SSDs. 
Limited hardware resources (such as main memory capacity) and inefficient random accesses prevent a straightforward implementation of the \RSGA pipeline inside storage. \textbf{Our goal} in this work is to leverage ISP capabilities in a careful way to accelerate \RSGA by alleviating the I/O overheads and accelerating key computational steps. 

\section{\revC{\proposal Key Idea}}

The core design idea for \proposal is to enable multiple Processing-In-Memory paradigms within the SSD and leverage \revC{the high SSD-internal} flash channel bandwidth to create a highly-parallel heterogeneous computing environment for \RSGA inside the storage system. 
Our optimization strategy consists of two key components:
First, we propose targeted software modifications on existing \RSGA pipelines that take into account SSD limitations and parallelization capabilities while maintaining accuracy. 
Second, we provide specialized near-memory computation units within the SSD for individual computational steps of the \RSGA pipeline and orchestrate the data flow between them. 
We leverage two computational approaches within the SSD: 
(i) ~Processing-Using-DRAM, which exploits the analog properties of DRAM arrays in SSD to perform massively parallel in-memory operations with minimal data movement overhead.  
(ii)~Processing-Near-DRAM, which adds lightweight compute logic \emph{close} to the SSD's internal DRAM, either \textit{near the DRAM subarrays} or \textit{inside the SSD controller}, tailored to the demands of each \RSGA step.

\section{\proposal \revB{ Genome Analysis} Workflow}
\label{sec:rawgains-sw}

\proposal implements a genome analysis workflow based on the state-of-the-art \RSGA approach presented in Fig.~\ref{fig:rawhash-flow}. 
The scope of \proposal's software modifications is to reduce both computational workload and intermediate storage requirements, resulting in a version of \RSGA that is optimized for efficient in-storage execution.

\subsection{Filtering Techniques}
\label{sec:mech-filt}

We adopt two distinct filtering techniques to reduce the load on the computationally intensive and resource-demanding chaining step.  

\noindent\textbf{Frequency Filters.} First, we leverage \textit{frequency filters}~\cite{firtina_rawhash2_2023,li_minimap2_2018} to only examine unique, meaningful matches \revB{between signal queries, e.g., seeds, and reference genomes}. 
Frequency filters are applied to the hash values \revB{created by multiple seeds (\S\ref{sec:BG-RSGA})}, and discard seeds that appear within the reference genome above a predefined threshold frequency $(thresh\_freq.)$.

\noindent\textbf{Seed-and-Vote Filtering.} Second, we adopt the \textit{seed-and-vote} filtering technique~\cite{liu_fast_2016,liao_subread_2013} to discard anchors unlikely to generate a correct alignment.
As described in Section~\ref{sec:background}, we partition the reference genome into windows, and anchors vote for windows that contain exact matches. 
A window with a high number of votes is more likely to contain the correct alignment.
Only windows receiving a number of votes above $(thresh\_voting)$ are retained for further processing.
This threshold is selected to balance accuracy (measured via F1-score) and performance, ensuring sufficient anchors are preserved for sensitivity, while discarding redundant matches to reduce workload.
This is the first work to apply the seed-and-vote technique to raw signals. 
For raw signals, this process is particularly challenging because reads and references, when converted to events, can include noise. 
To address this, we apply the seed-and-vote technique \textit{after the quantization and hash-table query steps}, to preserve  accuracy.

Based on the size and characteristics of the target genome, \revC{parameter values} for both filtering techniques, \revC{i.e., $thresh\_freq$, $thresh\_voting$, 
%the maximum seed frequency used for frequency filtering 
and the window size %and threshold on the minimum number of matching seed votes 
for seed-and-vote filtering} may vary. 
To ensure robust performance across a wide range of datasets, we perform an offline parameter space exploration to tune the parameter values and achieve a fair trade-off between accuracy and performance of the analysis.
Our exploration space is defined by the tuple ($thresh\_freq$, {$thresh\_voting$, $voting\_window$}).
We test different configurations on a subset of \revB{each} dataset (0.5-2\%) and observe that genomes with similar properties (e.g., size or complexity) consistently benefit from the same parameter configurations.
Small genomes yield the \revB{best trade-off between accuracy and performance} across different datasets for \revB{values} of $(2000,5,256)$ and large genomes for $(20000,2,256)$. 
Although the values cover a representative set of diverse genomes, they are easily reconfigurable for new genome types. 
\revC{The parameter exploration is performed only once offline and therefore does not impact the end-to-end runtime and energy.}

\subsection{Arithmetic Conversion Techniques}\label{sec:5.2}

We improve the utilization of the internal flash-channel bandwidth available within the SSD by using \revB{arithmetic conversion techniques}. The key idea of this optimization is to convert \textbf{floating-point} values to \textbf{fixed-point} and benefit from reduced storage requirements (i.e., mostly reduced bit-width from 64 or 32 bits to 16 bits) for intermediate data, as well as enable resource-efficient and less time-consuming fixed-point operations. We perform an experimental analysis \revB{at} software level and evaluate the accuracy achieved for \revB{fixed-point} arithmetic using 32, 16 and 8 \revB{bits}. The use of 16 \revB{bits} \revB{leads to small accuracy loss} compared to floating point and significant resource utilization savings.

Our goal is to maximize savings by applying \revB{arithmetic} conversion as early as possible in the pipeline.
However, adopting fixed-point arithmetic at the beginning of the pipeline is challenging due to the noise of raw signals, i.e., leveraging \revB{fewer} bits for raw signals interferes with subsequent signal-to-event conversion and quantization leveraged in typical \RSGA pipelines~\cite{firtina_rawhash2_2023}, leading to \revB{much lower} accuracy.
Applying \revC{early quantization, i.e., applying quantization directly on the raw signal} before signal-to-event conversion, alleviates this challenge. It increases stability against possible noise and facilitates the adoption of fixed-point arithmetic.
Unlike previous works~\cite{firtina_rawhash2_2023}, our \revB{workflow} first applies \revB{quantization}, followed by converting floating-point to fixed-point arithmetic, and then executes the signal-to-event conversion.
We show the accuracy results of our implementation for both fixed- and floating-point \revB{in \S\ref{sec:eval}.}
%\todo{I want a figure with the new pipeline here but I am not sure if there is space. If it is important I could replace Fig.2}

\section{\proposal Architecture and System}\label{sec:mech-and-sys}

We propose \proposal, the first ISP system designed for accelerating \RSGA by reducing data movement overheads and leveraging highly parallel computation capabilities present inside modern \revC{storage systems}. 
We design \proposal as an end-to-end \isp system that expands the capabilities of \revC{state-of-the-art} SSDs and \revC{autonomously executes the \RSGA pipeline without host intervention}. 

\subsection{\proposal In-Storage Architecture}
\label{sec:mech-overview}

\subsubsection{\textbf{Overview}}
\begin{figure}[h]
  \centering
  \includegraphics[width=\columnwidth]{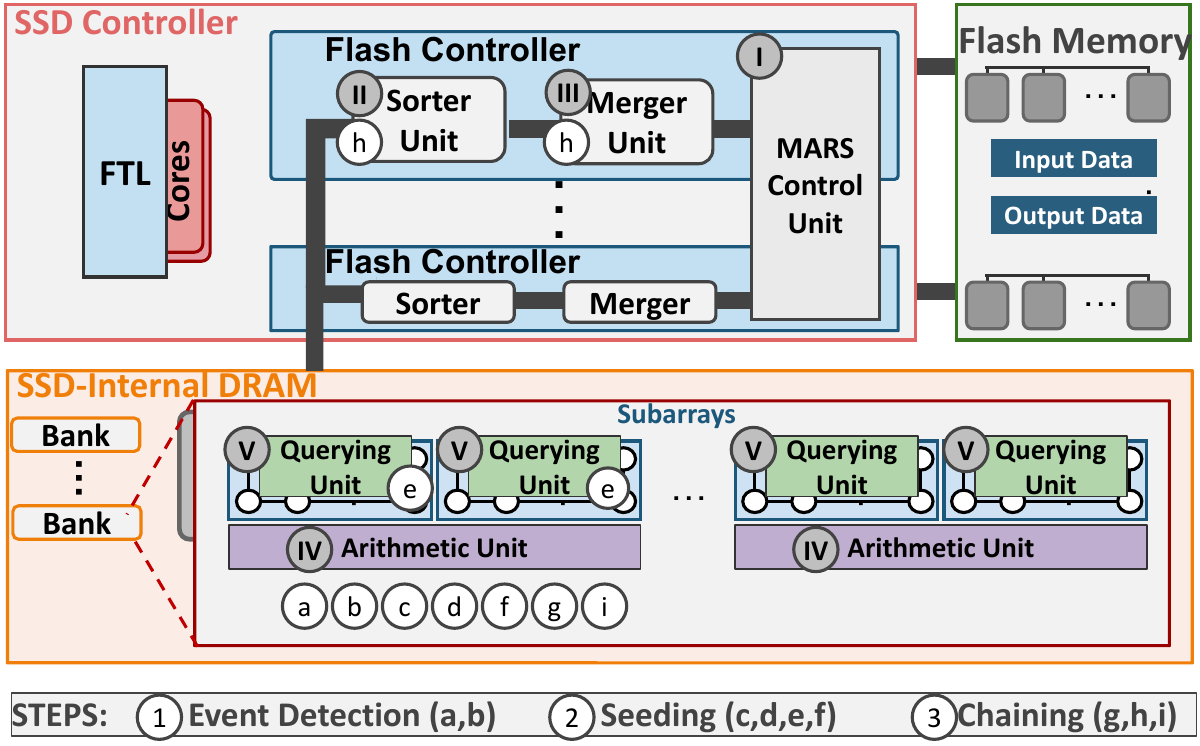}
  \caption{High-level overview of \proposal architecture.}
  \label{fig:overview-RG}
\end{figure}

Fig.~\ref{fig:overview-RG} shows a high-level overview of our system and the application flow.\footnote{\revC{To ease readability,} Fig.~\ref{fig:overview-RG} and \ref{fig:sorting} exclude control paths.} 
\proposal consists of five types of components: \textit{MARS Control Unit}~\circledgray{\footnotesize{I}}, \textit{Sorter Unit}~\circledgray{\footnotesize{II}}, \textit{Merger Unit}~\circledgray{\footnotesize{III}}, \textit{Arithmetic Unit}~\circledgray{\footnotesize{IV}} and \textit{Querying Unit}~\circledgray{\footnotesize{V}}.

\noindent\textbf{\revC{SSD Controller Components:}} \textit{MARS Control Unit}, \textit{Sorter Unit} and \textit{Merger Unit} are placed inside the SSD controller. 
\textit{MARS Control Unit}\revC{~\circledgray{\footnotesize{I}}} \revC{acts} as a Finite State Machine (FSM) that controls and coordinates the data flow between \proposal computation units.
\proposal \textit{Sorter Unit}\revC{~\circledgray{\footnotesize{II}}} (\S\ref{sec:pns}) is an accelerator that sorts sequences \revC{up to a predefined length.} 
The \textit{Merger Unit}\revC{~\circledgray{\footnotesize{III}}} (\S\ref{sec:pns}) efficiently combines short sorted sequences into longer ones.
\revC{Both units follow} the Processing-Near-DRAM approach, operating on data that originates from SSD-internal DRAM. \revC{One} Sorter and Merger pair is added per Flash Controller, \revC{adding up to} 8 instances.

\noindent\textbf{\revC{SSD-internal DRAM Components:}} The \textit{Arithmetic Unit}\revC{~\circledgray{\footnotesize{IV}}} and \textit{Querying Unit}\revC{~\circledgray{\footnotesize{V}}} are \revC{placed} inside the SSD-internal DRAM \revC{chips}.
The \textit{Arithmetic Unit}\revC{~\circledgray{\footnotesize{IV}}} (\S\ref{sec:pnm}) \revC{performs} arithmetic and logical operations. 
It \revC{leverages} the Processing-Near-DRAM approach: An Arithmetic Unit is placed at the edge of each pair of subarrays' peripheral logic, leading to 256 instances.
The \textit{Querying Unit}\revC{~\circledgray{\footnotesize{V}}} (\S\ref{sec:pum}) \revC{performs efficient hash-table lookups.} It leverages the Processing-Using-DRAM paradigm, i.e., exploits the analog \revC{operational} properties of the SSD-internal DRAM:  One \textit{Querying Unit} is placed per subarray, leading to 512 instances.

\vspace{3mm}
%\vspace{3mm}

\subsubsection{\textbf{Mapping Workflow to Compute Units}} 
We perform a detailed analysis of the \RSGA workflow to partition the \RSGA steps (i.e., event detection, seeding, chaining) into more fine-grained tasks such as arithmetic (e.g., addition, multiplication, division), querying and sorting operations.
The entire \RSGA workflow is described as a pipeline of these fine-grained tasks and each one is mapped to one of the available computation units (Arithmetic, Querying, Sorter, or Merger Unit) for efficient execution.
The MARS Control Unit encodes the pipeline steps and the order of their execution into a Finite State Machine and sequentially orchestrates them at runtime. 
While the pipeline sequence is predefined, actual computations in each step are triggered dynamically based on \revC{the availability of} inputs. 
Each compute unit is activated only when its inputs are available, ensuring resource efficiency and avoiding contention.

\subsubsection{\textbf{Control and Data Flow}}
Each step in \proposal's pipeline begins as soon as the previous \revC{one} finishes.
Before starting execution, the index and raw input data are distributed uniformly in terms of size across all SSD channels.
Data is transferred to the \proposal \textit{Arithmetic Units}~\circledgray{\footnotesize{IV}}, close to the SSD-internal DRAM subarrays. The  \textit{Arithmetic Units}~\circledgray{\footnotesize{IV}} perform the \textit{event detection} step~\circled{\footnotesize{1}} consisting of signal-to-event conversion~\circled{\footnotesize{1a}} and quantization~\circled{\footnotesize{1b}}, executed sequentially.
Next, as part of the seeding step~\circled{\footnotesize{2}}, the \textit{Arithmetic Units} execute the hash-value generation~\circled{\footnotesize{c}} and the frequency filter~\circled{\footnotesize{d}}. The filtered hash values are used for querying~\circled{\footnotesize{e}} the hash-table for seed hits inside the DRAM at the \textit{Querying Units}~\circledgray{\footnotesize{V}}.
The seed-and-vote filtering~\circled{\footnotesize{f}} step discards non-promising seed hits by leveraging once more the \textit{Arithmetic Units}~\circledgray{\footnotesize{IV}}.
During the chaining step~\circled{\footnotesize{3}}, the data is first bucketized~\circled{\footnotesize{g}} within the \textit{Arithmetic Units}~\circledgray{\footnotesize{IV}} and transferred to the \textit{Sorter}~\circledgray{\footnotesize{II}} and \textit{Merger Units}~\circledgray{\footnotesize{III}} inside the SSD controller for the sorting~\circled{\footnotesize{h}} step. 
%Sorting takes place in dedicated logic units near the flash controller based on a custom sort-and-merge technique (\circled{D}, \S\ref{sec:pns}).
The sorted data fragments are consolidated back in the SSD-internal DRAM for the final part of chaining, i.e., a dynamic programming-based algorithm~\circled{\footnotesize{3i}} implemented within the \textit{Arithmetic Units}~\circledgray{\footnotesize{IV}}.

%\vspace{-1.5mm}

\subsection{Event Detection Implementation}
\label{sec:pnm}

We map event detection (i.e., signal-to-event conversion and quantization \S\ref{sec:BG-RSGA}) to the \textit{Arithmetic Unit} as it mainly comprises additions and multiplications. 
\revC{One \textit{Arithmetic Unit} is placed next to two SSD-internal DRAM subarrays} to perform arithmetic operations close to the data and leverage the large subarray-level parallelism available within the DRAM.
\proposal is the first work to implement Processing-Near-DRAM inside the storage-internal DRAM.

\begin{figure}[h]
  \centering
  \includegraphics[width=0.9\columnwidth]{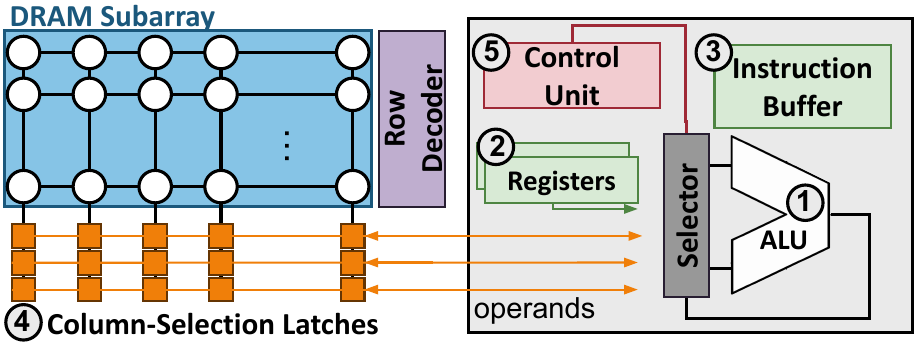}
  \caption{Overview of the \proposal Arithmetic Unit near a DRAM subarray.}
  \label{fig:pim-enabled-dram}
\end{figure}

\noindent\textbf{Arithmetic Unit Architecture and Mechanism.} Our design is inspired by a previous DRAM-based design, FULCRUM~\cite{Lenjani_2020_Fulcrum}. Fig. \ref{fig:pim-enabled-dram} illustrates the main components of the design.
A single-word \textit{ALU}~\circled{\footnotesize{1}} is placed next to a DRAM subarray and performs addition, comparison, multiplication, and bitwise operations. \textit{Registers}~\circled{\footnotesize{2}} are placed near the ALU to store intermediate results.
A programmable \textit{Instruction Buffer}~\circled{\footnotesize{3}} stores pre-decoded information for potential instructions, i.e., different operands and branch outcomes. 
\textit{Column-Selection Latches}~\circled{\footnotesize{4}} are placed on each column of each subarray to enable \revC{sequential access to individual columns~\cite{Lenjani_2020_Fulcrum}.}
A \textit{Control Unit}~\circled{\footnotesize{5}} determines the order of instructions and location of next access to the memory array.
\vspace{2mm}
\vspace{2mm}

In order to map the operations of signal-to-event conversion and quantization to the Arithmetic Unit, we first break each of them \revC{down} into arithmetic, predicate-based and condition-based operations. 
We construct pre-decoded instructions for all potential branches of execution within these operations and store them in the programmable instruction buffer. 
Based on the outcome of the previous operation, the Control Unit (i) selects the next instruction from the Instruction Buffer and (ii) identifies the columns of the subarray that need to be accessed.
This ensures that the Column-Selection Latches either capture the correct input operands (read from the subarray) or hold the correct target values before writing them back to the subarray.

\subsection{Seeding Implementation}
\label{sec:pum}

The hash-value generation, frequency filter and seed-and-vote filtering steps in seeding comprise arithmetic operations and pairwise comparisons. 
To execute those operations efficiently, we use the \textit{Arithmetic Unit} described in Section \ref{sec:pnm}.
However, hash table querying presents unique challenges due to the hash table's large size and the frequent random memory accesses it requires.
To address this, we implement the hash-querying mechanism inside SSD-internal DRAM leveraging Processing-Using-Memory, in particular the pLUTo~\cite{Ferreira2022pLUTo} approach. 
This method exploits DRAM's high storage density to enable massively parallel storage and querying of lookup tables (LUTs), ensuring efficient and scalable operations.

\begin{figure}[h]
  \centering
  \includegraphics[width=\columnwidth]{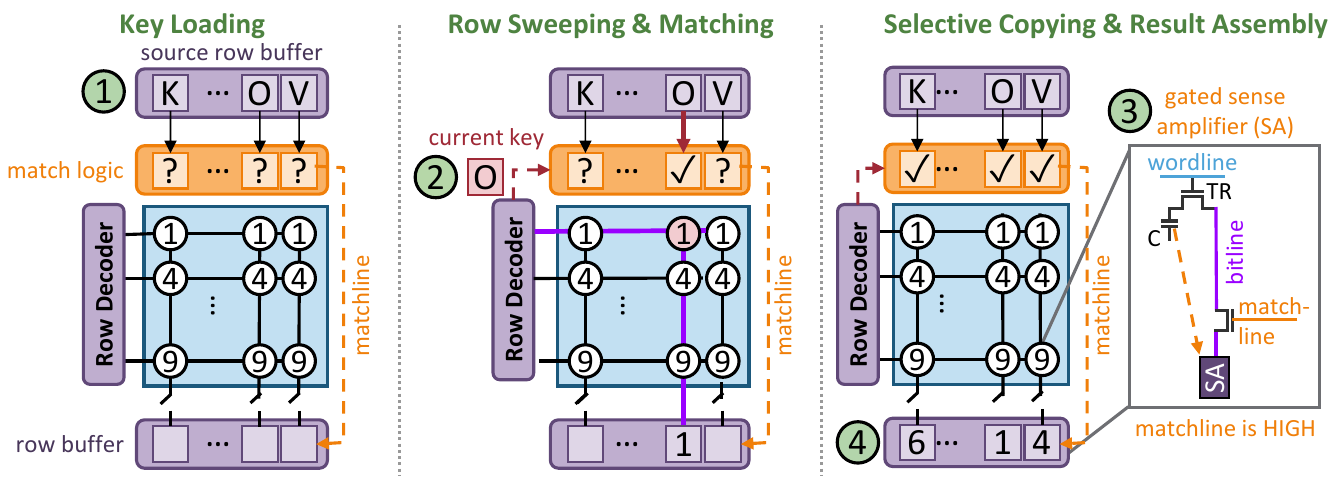}
  \caption{Overview of our hash table query mechanism in the SSD-internal DRAM.}
  \label{fig:pluto}
\end{figure}

\noindent\textbf{Querying Unit Architecture and Mechanism.} Fig.~\ref{fig:pluto} shows the architecture and step-by-step control flow of the Querying Unit. 
The hash table is stored in the \revC{SSD-internal} DRAM and is queried by subsequently activating DRAM rows using custom match logic and gated sense amplifiers (SA)~\cite{Ferreira2022pLUTo} (highlighted in orange). 
The custom match logic, located adjacent to the row buffer, uses comparators to compare the currently activated row index against the key values loaded into the source row buffer. A matchline is implemented as part of the custom match logic to enable the gated SA to selectively copy the corresponding value into the output buffer, when a match is detected.
A single query proceeds in four steps.
~\circled{1} \textbf{Key Loading.} The source row buffer is populated with the input keys (e.g., in Fig.~\ref{fig:pluto}: random values K, O, V). 
~\circled{2} \textbf{Row Sweeping \& Matching.} DRAM rows containing candidate hash entries are sequentially activated. 
For each row, the match logic compares the row index to the loaded keys. If a match is detected, the corresponding matchline is asserted.
~\circled{3} \textbf{Selective Copying.} The gated sense amplifiers sense and copy only those values in the currently activated row that correspond to matched keys.
~\circled{4} \textbf{Result Assembly.} The matched hash values (e.g., in Fig.~\ref{fig:pluto}: 6, 1, 4) are assembled in the row buffer.

\vspace{3mm}
\vspace{3mm}
 
If the DRAM size allows it, we store several copies of the hash table in the computation-enhanced subarrays to query multiple values in parallel.
If the genome index exceeds DRAM capacity, MARS adopts a partitioning strategy: large indexes (e.g., 52 GB \revC{for the human genome in D5}) are divided into smaller regions (e.g.,  2.6 GB), which are loaded into the SSD DRAM and queried sequentially. To minimize performance impact, MARS overlaps computation with data loading, effectively hiding the data movement latency.

\subsection{Chaining Implementation}
\label{sec:pns}
Chaining (\S\ref{sec:BG-RSGA}) consists of a sorting step (i.e., to sort seed positions) and a dynamic programming algorithm to extend chains from sorted seeds. 
While the dynamic programming part, based on additions and min operations, is efficiently handled by our near-DRAM \textit{Arithmetic Unit}, sorting large sets of seeds directly near DRAM would be either slow or require substantial area due to custom comparator logic.
Instead, we implement a highly parallel, custom sorter design inside the storage controller and benefit from increased scalability provided by the available SSD controller resources.

\noindent\textbf{Key Idea.} The main implementation challenge is efficiently sorting input sequences of variable length with high throughput and minimal area overhead.
We address this challenge by designing a resource-efficient hierarchical mechanism consisting of (1) a \textit{Sorter Unit} that processes input sequences of up to 128 elements
%using a fixed set of compare-and-swap (C\&S) units 
and (2) a \textit{Merger Unit} that combines smaller sorted subsequences into larger sorted outputs, enabling scalability beyond 128 elements.
%Our architecture is based on the bitonic sorter and merger~\cite{song2016parallel,samardzic2020bonsai,batcher1968sorting}, to benefit from their inherent parallelism and hardware-friendly structure and operations.

\noindent\textbf{Sorter and Merger Unit Architecture.} \proposal's Sorter and Merger Unit is based on the bitonic sorter and merger, respectively ~\cite{song2016parallel,samardzic2020bonsai,batcher1968sorting}, to benefit from their inherent parallelism and hardware-friendly structure and operations.
%The bitonic sorter uses a fixed set of compare-and-swap (C\&S) units, to recursively reorders elements to produce a fully sorted subsequence up to 128-elements long.
%For larger inputs, the high-throughput merger takes over. It processes two sorted input streams in a continuous, feedback-free pipeline without intermediate buffering, simplifying control and maximizing throughput.
Sorter and merger units are throughput-matched to prevent pipeline stalls and are sized to balance area efficiency with maximum utilization of the available internal SSD bandwidth.

\begin{figure}[h]
  \centering
  \includegraphics[width=0.95\columnwidth]{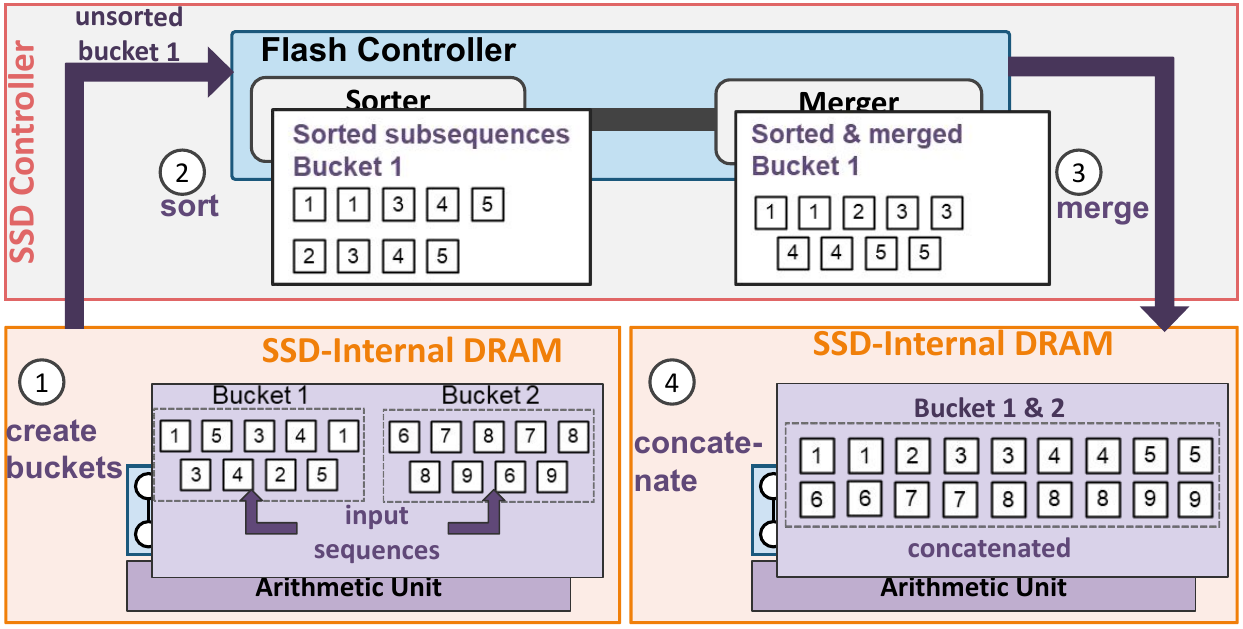}
  \caption{Simplified \revC{overview} of our Sort-and-Merge workflow.}
  \label{fig:sorting}
\end{figure}

\noindent\textbf{Mechanism.}
\proposal's Control Unit manages the sorting and merging process, including data movement between the SSD-internal DRAM and the \revC{Sorter and Merger Units. Fig.~\ref{fig:sorting} shows the  Sort-and-Merge mechanism flow.}

\vspace{3mm}
\vspace{3mm}

As shown in Fig.~\ref{fig:sorting}, ~\circled{1} \revC{the Contol Unit groups} unsorted seeds stored in SSD-internal DRAM into eight buckets, with each bucket corresponding to a non-overlapping region of the genome. 
~\circled{2} \revC{It transfers each} bucket to one of eight parallel Sorter–Merger units located near the storage controller. 
Each \revC{Sorter Unit splits its assigned} bucket into smaller subsequences, i.e., shorter than \revC{or equal to} 128 elements, and \revC{sorts them} locally using bitonic \revC{sorting}.
~\circled{3} \revC{If a bucket contains longer sequences, the Sorter Unit forwards the sorted subsequences to the Merger Unit.
The Merger Unit then merges these short, sorted input sequences into a longer fully sorted sequence using a streaming, one-pass merge strategy with no intermediate buffering or feedback.}
This design enables continuous, one-pass merging with low control complexity and high throughput, especially for long or variable-length inputs.
~\circled{4} \revC{The Control Unit writes the sorted outputs back to the SSD-internal DRAM.}
Since buckets are non-overlapping, they can be directly concatenated without further merging.
If local registers near the Merger Units are insufficient, \revC{the Control Unit temporarily buffers intermediate results in DRAM.} 
The final sorted sequences are subsequently consumed by the dynamic programming stage of chaining.

\subsection{System Integration}
\label{sec:mech-system}

\proposal is integrated into a \revC{modern} SSD with two \revC{different} modes of operation: \textit{conventional} and \textit{accelerator}. In \textit{conventional mode}, the SSD \revC{operates as a storage device only}. In \textit{accelerator mode}, a MARS-enabled SSD only performs \RSGA.  
This dual-mode of operation is feasible through \revC{small} changes to the Flash Translation Layer (FTL).

\noindent\textbf{\proposal FTL and Data Placement.} 
\revC{At} the beginning of \textit{accelerator} mode, \revC{the Control Unit flushes} all metadata essential to the \textit{conventional} mode
(e.g., the page status table, block read counts, logical-to-physical (L2P) mapping etc.) to the flash storage.
\proposal leverages the access pattern of the \RSGA workflow to \revC{apply} a storage-efficient custom logical-to-physical (L2P) mapping for the \textit{accelerator} mode.
Since the access pattern of the genome index and reference is sequential, \revC{data is placed} on the flash chips in a log-structured manner. 
\revC{The Control Unit then accesses the data in a sequential manner from the starting logical page address (LPA) and reads} across channels in a round robin manner.
Thus, \revC{this design allows to} keep a small mapping data structure consisting of: (1)~the mapping between the starting LPA and the physical page address (PPA), (2)~the database size, and (3)~a sequence of physical block addresses (PBAs) rather than the complete LPA-to-PPA mappings to store the genomic data.

\noindent\textbf{SSD Management Tasks.} 
\noindent\textit{Error Correction, Read Disturbance and Data Retention:} 
\revC{Since \proposal’s accelerators
operate within the SSD controller and the SSD-internal DRAM, all data is accessed by the Control Unit after ECC decoding~\cite{cai2017error,cai2017errors,zhao2013ldpc,tanakamaru2013error}.
\proposal effectively tackles read disturbance and data retention impact~\cite{cai2017error,ha2015integrated,luo2015warm,luo2018improving,cai2015read,cai2015data} since:
(i)~The sequential access pattern of the \RSGA pipeline minimizes repeated reads to the same page within short intervals, reducing the likelihood of read disturbances~\cite{cai2015read,ha2015integrated}, 
(ii)~Commodity SSDs automatically apply data refresh policies that refresh pages once their read counts exceed predefined thresholds,
(iii)~The time interval between subsequent refreshes does not exceed the duration of \RSGA,
which is substantially shorter than the manufacturer-specified threshold for reliable retention age (e.g., one year~\cite{micron3dnandflyer}).}

\noindent\textit{Wear-leveling:} \revC{\proposal effectively mitigates the impact of writes on flash lifetime thanks} to two design choices: 
(i) \revC{Our design employs} an out-of-place write policy and selects new blocks for writing based on their age, thereby effectively reducing long-term degradation, and (ii) \revC{\textit{flash writes} are minimized as the Control Unit only writes the final read mapping results} from the SSD-internal DRAM to the flash memory at the end of the \RSGA workflow. \\
\textbf{Storage Interface Commands}. 
\proposal operates independently of the \revC{host} during \RSGA execution, using the FSM in the SSD controller. 
\revC{Our design introduces} two new NVMe commands, \revC{i.e., standardized interfaces used by the host to communicate with SSDs,} for the host to support MARS execution: (i) \emph{\proposal{}\_Init} initiates the RSGA analysis and signals the SSD to switch from the conventional into the accelerator mode, (ii) \emph{\proposal{}\_Write} command updates both the \proposal FTL and regular FTL at the end of the application when the read mapping results are written from the SSD-internal DRAM to flash cells.

\section{Evaluation Methodology}
\label{sec:Methodology}

\noindent\textbf{Evaluated Systems.} 
We evaluate \proposal by comparing it against state-of-the-art \RSGA and conventional basecalling-based read-mapping systems in terms of accuracy, performance and energy. 
As a baseline for \RSGA-based read mapping, we select state-of-the-art \rhtwo~\cite{firtina_rawhash2_2023}, which offers a better accuracy-throughput trade-off compared to prior \RSGA tools \revC{and techniques}, including \rh~\cite{firtina_rawhash_2023}, Sigmap~\cite{zhang_sigmap_2021} and UNCALLED~\cite{kovaka_uncalled_2021}. 

We evaluate the following systems:
(1) \textbf{\bc}: a baseline pipeline for basecalling-based read mapping comprised of GPU-based basecaller \textit{Dorado}~\cite{dorado} and \textit{minimap2}~\cite{li_minimap2_2018} read-mapping tool (Version 2.24-r1122).
To simulate a real-time setting, we assume the basecaller processes raw signal chunks incrementally as they are generated by the sequencer rather than waiting for the full completion of each read's raw signal.
(2) \textbf{\rhCPU}: \rhtwo~\cite{firtina_rawhash2_2023} \RSGA-based read mapping baseline running on a state-of-the-art server-grade CPU~\cite{cpu}.
(3) \textbf{\rgCPUFP}: \proposal executed on CPU using floating-point arithmetic and the filtering optimizations presented in Section ~\ref{sec:rawgains-sw}.
(4) \textbf{\rgCPU}: \proposal executed on CPU using both fixed-point arithmetic and filtering optimizations.
(5) \textbf{\rg}: our proposed in-storage design of \proposal \revC{ using fixed-point arithmetic}, implemented as described in Section~\ref{sec:mech-and-sys}.
(6) \textbf{\rgEXT}: a variant of \proposal that \revC{add all computation units outside (external to) the SSD}. Sorting is offloaded to a near-CPU ASIC based on our custom design, while arithmetic and hash querying operations are executed in DRAM-based PIM units~\cite{Lenjani_2020_Fulcrum,Ferreira2022pLUTo}.
This configuration represents a PIM-only system that avoids any in-storage computation and serves as a comparison point to evaluate the benefits of tightly integrated compute within the storage hierarchy.
(7) \textbf{\rgSIMDRAM}: a \proposal variant that replaces the Processing-Near-DRAM-based Arithmetic Unit with a SIMDRAM-based~\cite{hajinazar2020simdram} Arithmetic Unit.
(8) \textbf{\genpip}~\cite{mao_genpip_2022}: a state-of-the-art hardware-accelerated, basecalling-based read mapping pipeline combining non-volatile memory (NVM)-based \PIM with algorithmic optimizations
(9) \textbf{MS-SmartSSD}~\cite{lee2020smartssd}: \revC{an existing system~\cite{smartssd_ug1382_2021} which directly connects an FPGA with the SSD via an external 3 GB/s link~\cite{wang2024ndsearch}. We map \proposal's Sorter and Merger Logic Units to the FPGA (300 MHz clock frequency ~\cite{smartssd_ug1382_2021}) and our PIM-components (\S\ref{sec:pnm},\ref{sec:pum}) in the SSD-internal DRAM}.

\noindent\textbf{CPU and GPU Configurations.} For the CPU-based systems, we use a high-end server with two 64-core AMD EPYC 7742 CPUs~\cite{cpu}, 1TB of DDR4 DRAM~\cite{ddr4} and a performance-optimized SSD~\cite{samsungPM1735_2020} connected \revB{to the CPU} via a PCIe4 interface~\cite{PCIE4}. 
For the \bc system, the basecalling step (\textit{Dorado}~\cite{dorado}) runs on an NVIDIA RTX A6000 GPU~\cite{gpu}.
All software tools support multi-threaded processing where each raw signal \revB{sequence} is handled by a separate thread. 
We run all tools with the best-performing configuration of 128 threads to compare against our system.

\noindent\textbf{SSD and DRAM Configurations.} %For \revC{our design \proposal and the \rgSIMDRAM comparison point using a Processing-In-Storage design}
\revC{To evaluate \proposal and \rgSIMDRAM}, we consider a performance-optimized SSD with internal LPDDR4 DRAM~\cite{ddr4} (Table~\ref{tab:setup}).  
Since accelerators and compute units operate sequentially, we simulate each component individually, including the data movement between them.
For DRAM-based components (i.e., \textit{Arithmetic} and \textit{Querying} Units), we use timing parameters extracted from the LPDDR4 DRAM model in CACTI7~\cite{CACTI7}.
We assume that single-word ALUs embedded in SSD-internal DRAM operate at 164 MHz.
For SSD components we use MQSim~\cite{tavakkol2018mqsim,mqsim_github}, a widely-adopted simulator for modern SSDs. 
Our \textit{Sorter} and \textit{Merger} Unit are implemented in Verilog HDL and synthesized using Synopsys Design Compiler~\cite{synopsysdc} at 1 GHz to obtain timing, area, and energy results.
\revC{We model data movement overheads by calculating the transfer latency between each computing and storage element based on the size of the data to be transferred and the available bandwidth between components.}
We combine the simulation results from DRAM and SSD simulators, the Verilog synthesis and the data movement overheads to evaluate the end-to-end performance of \proposal.
%Each step of the \RSGA pipeline is simulated in the most appropriate domain and we integrate the results by considering data movement overheads to estimate the overall system behavior.}
\vspace{4mm}
\begin{table}[h]
%\captionsetup{font=scriptsize}
\centering
\caption{\revB{Simulation configuration of our design.}}
\resizebox{\linewidth}{!}{\Huge
\begin{tabular}{@{}clrrrr@{}}\toprule
\textbf{Component}& \textbf{Detailed Configuration} \\\midrule
SSD & NVMe, PCIe 4.0, PCIe lane BW: 1.2 GB/s, TLC,\\
&8 channels, 8 chips/channel, tDMA: 16$\mu$s, tR (TLC): 22.5$\mu$s, \\
&flash channel BW: 1 GB/s, 4 ARM Cortex R7       \\\midrule
SSD-Internal DRAM    & 4 GB LPDDR4 DRAM, 16 banks, 512 subarrays, \\&256 rows/subarray, row size: 2048 bytes      \\\midrule
\revB{Sorter and Merger Unit}    & Frequency: 1 GHz  \\\midrule
\revB{Arithmetic Unit}      & Frequency: 164 MHz       \\\bottomrule
\end{tabular}}
\label{tab:setup}
\end{table}
\vspace{4mm}

\noindent\textbf{Datasets.}%To demonstrate the applicability of \proposal across diverse genomic contexts, 
We evaluate \proposal on five real-world datasets from different organisms, covering a wide range of genome sizes. 
Table~\ref{tab:datsets} summarizes the dataset characteristics~\cite{dataset-D1,dataset-D2,dataset-D3,dataset-D4,dataset-D5-FAB,dataset-D5-WGS} and reference genomes~\cite{ref-gen-D1,ref-gen-D2,ref-gen-D3,ref-gen-D4,Miga2020XChromosome,Rhie2022YChromosome}, all obtained from public repositories. We use the \textit{fast5} file format for our input data and assume that the data is already correctly placed, i.e. sequentially and evenly distributed across all SSD channels, for all evaluated systems.
\vspace{4mm}

\begin{table}[h]
%\captionsetup{font=scriptsize}
\centering
\caption{\revB{Details of datasets used in our evaluation.}}
\resizebox{\linewidth}{!}{\Huge
\begin{tabular}{@{}clrrrr@{}}\toprule
& \textbf{Organism} & \textbf{Reads (\#)} & \textbf{Bases (\#)}              & \textbf{Genome Size (bp)} & \textbf{Dataset Size} \\\midrule
D1 & \emph{SARS-CoV-2}   & 1,382,016      & 594 M                   & 29,903 & 11 GB\\\midrule
D2 & \emph{E. coli}      & 353,317        & 2,365 M                 & 5 M  & 27 GB\\\midrule
D3 & \emph{Yeast}        & 49,989         & 380 M                   & 12 M & 39 GB\\\midrule
D4 & \emph{Green Algae}  & 29,933         & 609 M                   & 111 M & 74 GB\\\midrule
D5 & \emph{Human HG001}  & 269,507        & 1,584 M                 & 3,117 M & 39 GB\\\bottomrule
% \multicolumn{7}{l}{For relative abundance estimation and contamination analysis, we combine the reads or}\\
% \multicolumn{7}{l}{reference genomes from the datasets we show in the read mapping section. We use the}\\
\end{tabular}}
\label{tab:datsets}
\end{table}

\vspace{4mm}

\section{Evaluation}
\label{sec:eval}

\subsection{Accuracy Analysis}
\label{sec:eval-accuracy}

We evaluate \textbf{\rhCPU}, \textbf{\rgCPU} and \textbf{\rgCPUFP} accuracy based on the ground truth generated by basecalling reads with Dorado~\cite{dorado} and mapping the generated \textit{basecalled} reads to the reference genome using minimap2~\cite{li_minimap2_2018}.
\revC{All hardware systems implement \rgCPUFP workflow and thus achieve the same accuracy.}
We use UNCALLED pafstats~\cite{kovaka_uncalled_2021} tool to identify true positives (\texttt{TP}: correct mappings), false positives (\texttt{FP}: incorrect mappings), and false negatives (\texttt{FN}: unmapped reads that are mapped in the ground truth)  based on the mapping position distance from the respective ground truth. 
Using these values, we calculate precision (\texttt{P = TP/(TP+FP)}), recall (\texttt{R = TP/(TP+FN)}), and the F\textsubscript{1} score (\texttt{F\textsubscript{1} = 2x(PxR)/(P+R)}).

We make two observations based on the accuracy results reported in \ifnum\todos=1\todo{Cannot place Table 3 after it is referenced.}\fi Table~\ref{tab:accuracy}. 
(1)~\rgCPU outperforms \rhCPU in terms of recall and F\textsubscript{1} score for all evaluated datasets, while maintaining on-par precision for small genomes and only a slight reduction in precision for larger genomes. 
This improvement is due to the integration of our two proposed filtering techniques \revC{(\S\ref{sec:mech-filt})} and early quantization \revC{(\S\ref{sec:5.2}) } which together eliminate ambiguous or redundant candidate matches, i.e., matches that are frequent, low-quality or non-specific, and allow the pipeline to focus on signal regions that are more likely to represent correct alignments.\ifnum\todos=1\todo{it has been said in background and filtering mechanisms}\fi
(2)~The use of fixed-point and integer operations instead of floating-point operations only minimally decreases accuracy for all datasets.

\subsection{\kon{Performance Analysis}}
\label{sec:eval-latency}

\begin{figure}[H]
  \centering
  \includegraphics[width=\columnwidth]{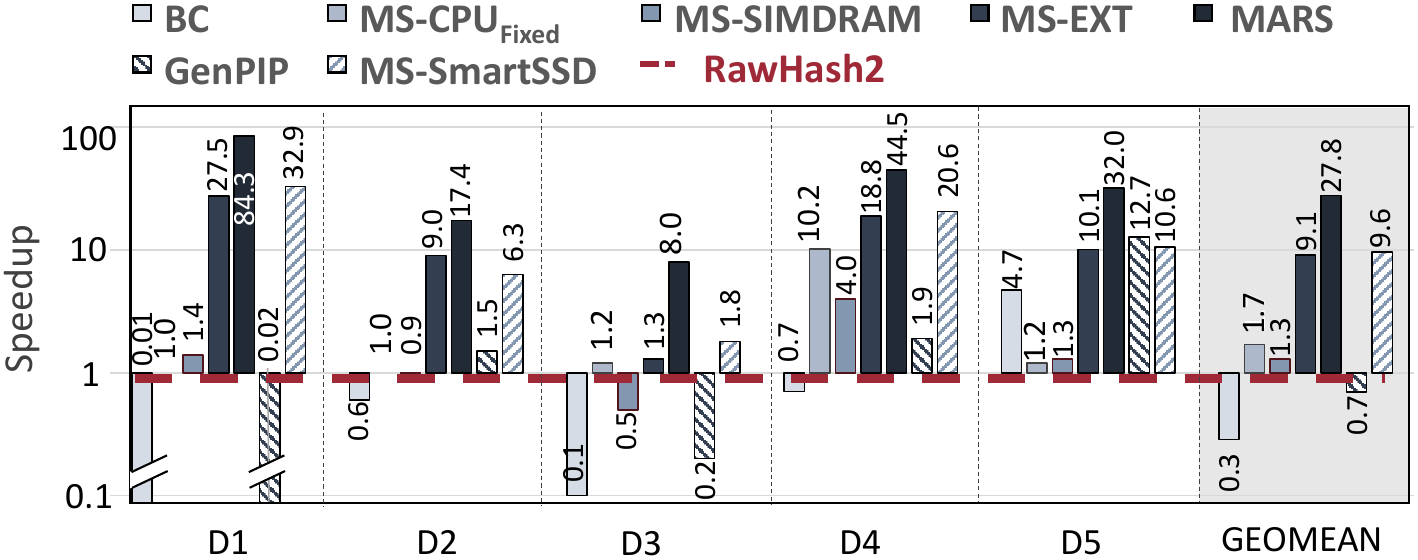}
  \caption{End-to-end execution time speedup of each system over \rhCPU.}
  \label{fig:latency-HW}
\end{figure}

We evaluate the performance of all \revC{seven} systems \revC{described} in \S\ref{sec:Methodology} leveraging the five diverse datasets of Table~\ref{tab:datsets}.
Fig.~\ref{fig:latency-HW} shows the execution time speedup achieved by each evaluated system over CPU-based \rhtwo, \rhCPU.
We make three observations.
First, ~\rg outperforms \emph{all} other baselines across \emph{all} datasets.
Compared to the GPU-accelerated basecalling based pipeline \bc, \rg delivers a speedup of 93$\times$ on average across all five datasets with larger speedups for smaller genomes. 
%This is due to \proposal minimizing the number of computations through filtering and performing read mapping directly on raw signals, eliminating basecalling, reducing data movement, and enabling highly parallel in-storage execution.
This is \revC{because \proposal (i)~eliminates basecalling, (ii)~applies filtering mechanisms, (iii)~reduces data movement, and (iv)~enables highly parallel in-storage execution.}

Second, \rg outperforms all prior hardware-accelerated solutions: \rgEXT, \rgSIMDRAM, \gp, and MS-SmartSSD.
Specifically, \revC{\rg improves} performance by 3.1$\times$ on average \revC{over} \rgEXT, which adopts PIM solutions (\proposal-based ASIC and PIM accelerator) outside the storage. 
This comparison point shows that \rgEXT fails to fundamentally solve the I/O data movement overhead problem and highlights the importance and need for in-storage processing for \RSGA. 
MS-SmartSSD performs worse than \rg, due to its limited 3 GB/s bandwidth between SSD and FPGA~\cite{lee2020smartssd}, which restricts the use of internal SSD bandwidth between flash and storage controller, fully utilized by \rg.
While \rgSIMDRAM addresses I/O overhead through in-storage computation, its use of bit-serial operations for arithmetic (e.g., multiplication, division) results in \revC{execution time 21.4$\times$ slower than \rg}.
\rg is the only design that both eliminates the I/O bottleneck and meets the computational demands of \RSGA acceleration.
%, outperforming all prior hardware-accelerated systems across all datasets (e.g., \gp by 2.9$\times$ on average).

Third, our algorithmic improvements alone (\rgCPU), i.e., without leveraging \ISP capabilities, provide a considerable speedup of 1.2-10.2$\times$ over \rhCPU for medium- to large-sized genomes (i.e., D3-D5) and on-par \revC{performance} for small genomes (i.e., D1,D2). 
This demonstrates the effectiveness of our software optimizations, including filtering, in reducing the computational load, particularly during chaining.

\noindent\revC{\textbf{Throughput evaluation}.}
We compare ~\rg's throughput with the  throughput of a single sequencer, which \revC{is} 450 bases per second (i.e., 4000 - 5000 samples per second)~\cite{Shan_rapid_2018}. 
As Table~\ref{tab:throughput} \revC{shows},
~\rg's throughput is substantially higher than 450 bp/sec for all datasets.
In fact, ~\rg outperforms the real-time analysis requirement of a full MinION sequencer~\cite{jain_minion_2016}, which processes data at 230,400 bp/s, by 46$\times$ on average across all datasets (between 1.2$\times$ for large genomes (D5) to 202$\times$ for small genomes (D1)).

%\vspace{4mm}

\begin{table*}[h]
\centering
\caption{Mapping accuracy of three \RSGA pipelines compared to basecalling-based ground truth.}
\resizebox{\linewidth}{!}{\LARGE
\begin{tabular}{l||lll|lll|lll|lll|lll}
\hline
           & \multicolumn{3}{c}{\textbf{D1 \emph{SARS-CoV-2}}} & \multicolumn{3}{c}{\textbf{D2 \emph{E.coli}}} & \multicolumn{3}{c}{\textbf{D3 \emph{Yeast}}} & \multicolumn{3}{c}{\textbf{D4 \emph{Green Algae}}} & \multicolumn{3}{c}{\textbf{D5 \emph{Human HG001}}}  \\ \hline
           &  Prec.        & Recall       & $F_1$        &  Prec.       & Recall      & $F_1$       &  Prec.      & Recall     & $F_1$       &  Prec.        & Recall       & $F_1$         &  Prec.        & Recall       & $F_1$         \\\hline
%\textbf{\bc} & 0.9900 & 1.0000 & 0.9950 & 0.9981 & 1.0000 & 0.9990 & 0.9609 & 1.0000 & 0.9801 & 0.8697 & 1.0000 & 0.9303 & 0.3941 & 1.0000 & 0.5654 & &\textbf{BC}\\\hline
\textbf{\rhCPU}         & 0.9868     & 0.8735     & 0.9267    & 0.9573    & 0.9009    & 0.9282   & 0.9862   & 0.8412   & 0.9079   & 0.9691     & 0.7015     & 0.8139     & 0.8949     & 0.4054     & 0.5582   \\\cline{1-16}
\textbf{\rgCPU} & \betterRH{0.9917}     & \betterRH{0.9694}     & \betterRH{0.9803}    & \betterRH{0.9854}    & \betterRH{0.9574}    & \betterRH{0.9712}   & 0.9533   & \betterRH{0.9643}    & \betterRH{0.9588}   & 0.9125     & \betterRH{0.9166}      & \betterRH{0.9141}     & 0.8723     & \betterRH{0.6318}     & \betterRH{0.7300}     \\\cline{1-16}
\textbf{\rgCPUFP}         & \betterRH{0.9939}     & \betterRH{0.9796}     & \betterRH{0.9867}    & \betterRH{0.9893}    & \betterRH{0.9616}    & \betterRH{0.9753}   & 0.9551   & \betterRH{0.9655}   & \betterRH{0.9603}   & 0.9254     & \betterRH{0.9438}     & \betterRH{0.9354}     & 0.8763     & \betterRH{0.6729}     & \betterRH{0.7612}  \\\hline  
\end{tabular}
}
\label{tab:accuracy}
\end{table*}

\begin{table}[h]
\centering
\caption{Throughput of \rg. A single nanopore has a throughput of 450 bp/sec; an entire MinION sequencer achieves 230,400 bp/sec.}
\resizebox{\linewidth}{!}{\begin{tabular}{@{}crrrrr@{}}\toprule
                 & \textbf{D1} & \textbf{D2} & \textbf{D3}  & \textbf{D4} & \textbf{D5} \\\midrule
Throughput [bp/sec] & 46,655,128      & 5,274,148      & 1,202,660      & 1,277,764     & 286,728    \\\bottomrule
% \multicolumn{7}{l}{For relative abundance estimation and contamination analysis, we combine the reads or}\\
% \multicolumn{7}{l}{reference genomes from the datasets we show in the read mapping section. We use the}\\
\end{tabular}}
\label{tab:throughput}
\end{table}

%Since an entire MinION sequencer is equipped with 512 nanopores (i.e., ~230,400 bases per second), a single \proposal accelerator can provide the real-time analysis for an entire sequencer as its throughput is larger than the sequencer of an entire MinION throughput for all datasets. 
%(2)~\rg provides a better throughput improvement than \sqf for the D1 dataset. Specifically, \sqf~\cite{dunn_squigglefilter_2021} reports close to $114\times$ improvement over MinION sequencer while \proposal achieves almost $203\times$ higher throughput. 

\subsection{Energy Analysis}
\label{sec:eval-energy}
To demonstrate the energy benefits of \rg, we measure the energy consumption of all components (i.e., SSD, DRAM, CPU and if applicable GPU) involved in the respective systems. We use AMD µProf~\cite{AMDuProf} to measure the energy consumption for CPU-based systems, and the CACTI7~\cite{CACTI7} DDR4 model to estimate the power overheads on our \PIM-enabled DRAM design. We synthesize logic components with the Synopsys Design Compiler~\cite{synopsysdc} using a 65nm process node to estimate their power consumption.

\begin{figure}[h]
  \centering
  \includegraphics[width=1\columnwidth]{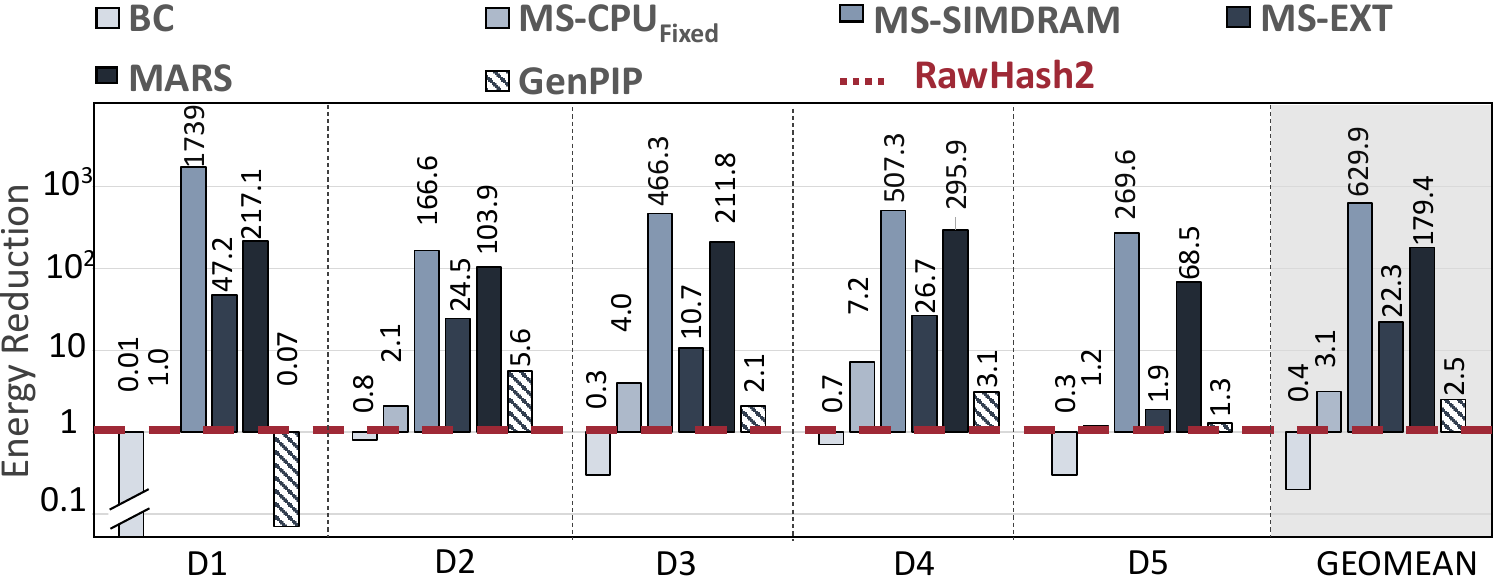}
  \caption{Energy reduction of each system compared to \rhCPU.}
  \label{fig:energy-HW}
\end{figure}

Fig.~\ref{fig:energy-HW} \ifnum\todos=1 \todo{Now it depicts how many times RH2 energy is reduced and is a "larger is better" axis. Also added some numbers.} \fi shows the end-to-end \revC{energy reduction achieved by all evaluated systems over \rhtwo (\rhCPU)}. We make three observations. 
(1)~All hardware-accelerated systems, i.e., \rg, \rgEXT, \rgSIMDRAM, \genpip \revC{ achieve greater energy reduction} compared to CPU-based setups, i.e., \bc and \rgCPU.
(2)~Only \rgSIMDRAM \revC{ yields higher energy reduction} compared to \rg \revC{ (by 3.5$\times$ on average across datasets)}, due to its simplified Arithmetic Unit based on bit-serial, in-memory execution. 
However, because of \rgSIMDRAM's significantly higher latency (\S\ref{sec:eval-latency}), \rg still provides a more favorable trade-off between latency and energy consumption.
(3)~\rgEXT \revC{ reduces energy by 22.3$\times$ as opposed to \rg's 79.4$\times$ reduction over \rhCPU}, due to high data movement from the storage to the host and accelerators and a greater reliance on the CPU for orchestration, which increases energy use on the host side.
\revC{Overall, \rg achieves the best energy consumption and performance trade-off among all designs}.

\vspace{2mm}

\subsection{Area Analysis}
\label{sec:eval-area}
\noindent\textit{\revC{SSD-internal DRAM overhead}.} We estimate the base area of our \PIM-enabled DRAM with CACTI7~\cite{CACTI7} to be 55.48 $mm^2$ in a 22nm technology. Each Arithmetic Unit occupies 0.0295 $mm^2$, leading to 7.56 $mm^2$~\cite{Lenjani_2020_Fulcrum,Lenjani_2022_supporting} total overhead for all 256 Arithmetic Units.
\revC{Each LUT-based Querying Unit occupies 0.018 $mm^2$,  leading to  9.22 $mm^2$~\cite{Ferreira2022pLUTo} for 512 instances}. 
The total DRAM overhead of our design, i.e., 16.78 $mm^2$, is low compared to the total SSD area available, \revC{i.e., at least 6400 $mm^2$ for our SSD configuration of 8 channels and 8 typical 100 $mm^2$ NAND flash chips per channel}.

\noindent\textit{\revC{SSD Controller Logic} overhead.} We estimate the area overhead of our logic components using Synopsys Design Compiler~\cite{synopsysdc} with UMC 65nm technology node~\cite{umc}. The area for the Sorter, Merger and Controller Unit is 0.78 $mm^2$, 0.14 $mm^2$ and 0.002 $mm^2$, respectively. Compared to a 14nm Intel Processor~\cite{wikichipcascade}, the Sorter and Merger introduce only 0.028\% area overhead (the area is 0.09 $mm^2$ when scaled to 14nm~\cite{STILLMAKER_2017_scaling})\ifnum\todos=1\todo{Omitted comparison with ARM}\fi. 
%This area overhead amounts to less than 25\% of three ARM Cortex R7~\cite{cortexr7}, multiple of which are deployed in the storage controller~\cite{MarvellBraveraSC5} of a PCIe-based SSD.
%Lastly, compared to the area introduced by the FPGA in the SmartSSD, our design only introduces an area overhead of 0.003\%~\cite{STILLMAKER_2017_scaling, wang2024ndsearch}.
%As our design provides significant performance speedups and high reductions in energy consumption, we conclude that this result is a good performance-energy-area trade-off.

\begin{table}[h]
\centering
\caption{Area analysis overview per component.}
\resizebox{\linewidth}{!}{\begin{tabular}{@{}llrrr@{}}\toprule
\textbf{Placement} & \textbf{Unit}  & \textbf{Instances} & \textbf{Area [$mm^2$]} &  \textbf{Area [$mm^2$]}\\
  \revC{in SSD}&  & \textbf{Number} & \textbf{per Unit} & \textbf{Total}\\\midrule
%\multicolumn{2}{l}{\textbf{Total for SSD}} & \textbf{7.37} \\\midrule
%& & &3 ARM Cortex R7\\
%\hline
%& &  & 0.028\% over Intel Processor 14nm\\\midrule
%& & & Intel Processor 14nm \\\midrule
\textbf{\revB{SSD-internal DRAM}} & \revB{Arithmetic} & 256  & 0.0295   & 7.56 \\\midrule
& \revB{Querying} & 512 & 0.018  & 9.22 \\\midrule
\hline
\textbf{\revB{SSD controller}} & \revB{Sorter}      & 8  & 0.78 & 6.24  \\\midrule
& \revB{Merger} & 8    & 0.14 & 1.12  \\\midrule
& \revB{Control} & 1 & 0.002 & 0.002 \\\midrule
%\multicolumn{2}{l}{\textbf{Total for SSD}} & \textbf{47.68}  \\\bottomrule
\end{tabular}}
\label{tab:area}
\end{table}
%\todo{maybe omit table, it is duplicate info. Add numbers in abstract,intro instead.}
\pagebreak
\subsection{Sensitivity to \revC{SSD-Internal DRAM Size}}
\label{sec:eval-sensitivity}

\begin{figure}[h]
  \centering
  \includegraphics[width=1\columnwidth]{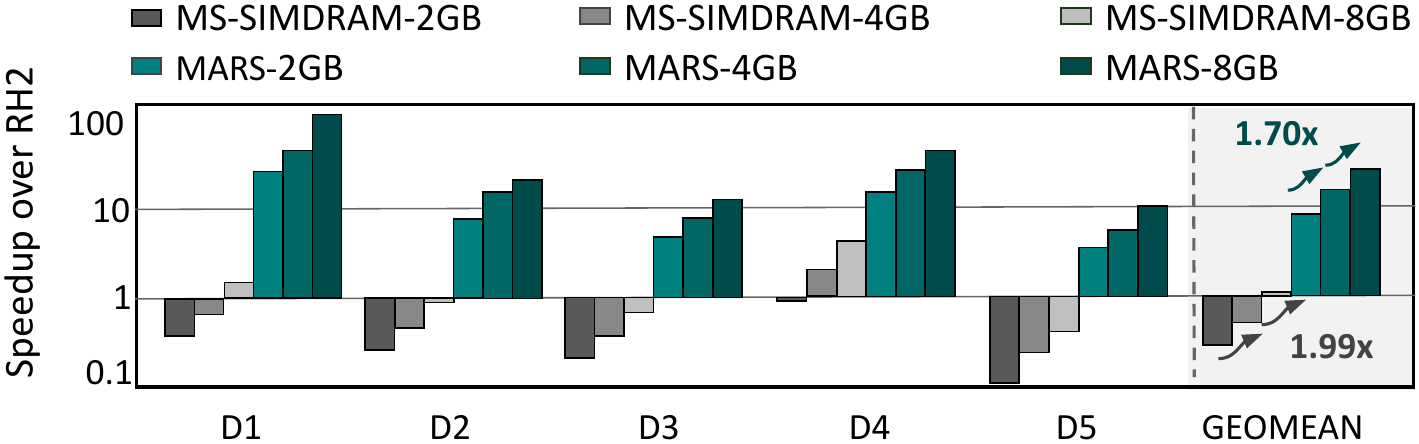}
  \caption{Sensitivity to SSD-internal DRAM size.}
  % \caption{Sensitivity analysis for the impact of internal DRAM size on \rgSIMDRAM and \rg speedup over \rhCPU.}
  \label{fig:sensitivity-DRAM}
\end{figure}

\revC{We perform a sensitivity analysis to examine the scalability of the two \ISP designs \rg and \rgSIMDRAM for different sizes of the SSD-internal DRAM, i.e., 2 GB, 4GB (base configuration) and 8 GB.
Fig.~\ref{fig:sensitivity-DRAM} shows %the speedup of \rg and \rgSIMDRAM over \rhCPU for different DRAM \revC{sizes}, i.e., 2 GB, 4GB (base configuration) and 8 GB.
that \rg's performance increases by 1.70x on average when we double the internal DRAM size, while \rgSIMDRAM's performance increases almost by 1.99x on average. Therefore, the proposed design scales well when increasing internal DRAM resources and is not bound by the internal bandwidth. \rgSIMDRAM's slightly better scaling indicates that increasing the DRAM capacity yields better results for PuM-based computations.} 
\vspace{4mm}

\section{Related Work}
\label{sec:rel-work}
%To our knowledge, \proposal is the \emph{first} end-to-end design for \RSGA scalable to large genomes. It enhances the capabilities of an SSD with data-centric compute primitives to optimally mitigate the I/O data movement bottleneck while providing large compute capabilities. 
%To our knowledge, \proposal is the first portable system for \RSGA that effectively addresses data movement and computation overheads of the end-to-end \RSGA pipeline for a wide range of genomic datasets.
%We have already extensively compared \proposal against existing state-of-the-art software implementations of \RSGA~\cite{firtina_rawhash_2023,zhang_sigmap_2021,kovaka_uncalled_2021}. 

\revC{To our knowledge, this is the first work to 1) enable in-storage acceleration of Raw Signal Genome Analysis and 2) combine the use of Processing-Near-Memory and Processing-Using-Memory inside the storage system. 
In this section, we briefly review prior work on hardware acceleration for genome analysis and \ISP.}

\noindent\textbf{Hardware Acceleration for \RSGA.}  Prior hardware acceleration works on \RSGA propose FPGA-based ~\cite{shih_haru_2023,10015864,sundaresan1992vlsi,samarasinghe2021energy} and GPU-based~\cite{Gamaarachchi2020_f5c,guo2019hardware,Bao2021Squigglenet,sadasivan_accelerated_2023} systems. Specifically, Squigglefilter~\cite{dunn_squigglefilter_2021} proposes an edge-GPU-based system for \RSGA that performs contamination analysis for small, viral genomes based on a 1D systolic array. HARU~\cite{shih_haru_2023} uses an MPSoC with an on-chip FPGA to accelerate \RSGA and f5c~\cite{Gamaarachchi2020_f5c} presents a GPU-based accelerator.
\revB{None of these systems 1) consider the impact of I/O data movement on end-to-end execution of \RSGA and 2) provide a system for genome analysis that is scalable to medium- and larger-sized genomes,} due to the use of costly dynamic time-warping alignment operations~\cite{dunn_squigglefilter_2021,lindegger2023rawalign,icdm10}. 
%\revB{\proposal effectively addresses both the I/O data movement bottleneck and the computational scalability limitations of prior \RSGA systems.}
\revB{A comparison of \proposal with SquiggleFilter and HARU is out of scope, as these works focus on performing read alignment for small, mostly viral genomes. \proposal can be integrated into these tools to help them quickly identify seed hits, thus avoid searching the entire genome and enable scaling to large genomes.}

\noindent\textbf{Hardware Acceleration for Genome Analysis.} Multiple prior works propose accelerator designs for basecalling-based genome analysis targeting basecalling and read mapping steps with different architectures like ASICs~\cite{turakhia2018darwin, fujiki2018genax, madhavan2014race}, GPUs~\cite{cheng2018bitmapper2, houtgast2018hardware, houtgast2017efficient, zeni2020logan, ahmed2019gasal2, nishimura2017accelerating, de2016cudalign, liu2015gswabe, liu2013cudasw++, wilton2015arioc, liu2009cudasw++, liu2010cudasw++, guo2019hardware}, FPGAs~\cite{fujiki2020seedex, banerjee2018asap, goyal2017ultra, chen2016spark, chen2014accelerating, chen2021high, fei2018fpgasw, waidyasooriya2015hardware, chen2015novel, rucci2018swifold, haghi2021fpga, li2021pipebsw, ham2020genesis, ham2021accelerating, wu2019fpga, liyanage2023efficient}, \ISP~\cite{mansouri2022genstore}, and \PIM~\cite{cali2020genasm, huangfu2018radar, khatamifard2021genvom, gupta2019rapid, li2021pim, angizi2019aligns, zokaee2018aligner,Zhang_2023_alignerD}. 
Basecalling accelerators~\cite{lou2020helix, lou2018brawl,shahroodi2023swordfish,ankit2019puma,singh2024rubicon,xu2021fastbonito,wu2020fpga,wu2022fpga,mao_genpip_2022} \revC{speed up} the translation of raw signals into nucleotide sequences, a step that is entirely bypassed by our \RSGA-based design.
Read mapping accelerators~\cite{cheng2018bitmapper2,houtgast2017efficient,chen2014accelerating,chen2021high,chen2015novel,khatamifard2021genvom,mansouri2022genstore,laguna2020seed,angizi2020pim,kaplan2018rassa,kaplan2020bioseal,doblas2023gmx,huang2023casa,Zuher_2023_DashCAM,liyanage2023efficient,gu2023gendp,haghi2023wfasic,pham2023accelerating,zhong2023asmcap,burchard2023space,gudur2021fpga} are \emph{not} applicable \revC{to} \RSGA as they do \emph{not} consider the noise within raw signals.

\noindent\textbf{\isp.} Prior works explore ISP through various approaches using (1)~Processing-Near-Flash memory by integrating processing capabilities into the SSD controller in a general-purpose~\cite{gu2016biscuit,kang2013enabeling,wang-eurosys-2019,acharya-asplos-1998,keeton-sigmod-1998, zou_assasin_2022} or application-specific way~\cite{mailthody2019deepstore,pei-tos-2019,Jun2018grafboost,do2013query, seshadri2014willow,kim2016storage,riedel-computer-2001,riedel-vldb-1998,Wang2024BeaconGNN}, (2)~Processing-using-Flash memory by exploiting the analog properties of flash memory~\cite{park2022flash, choi2020flash, han2019novel, shim2022gp3d, merrikh2017high, tseng2020memory, wang2018three, lue2019optimal,parabit_2021_gao} or by (3)~closely integrating SSDs with GPUs~\cite{cho2013xsd} or FPGAs~\cite{koo-micro-2017, jun2015bluedbm, torabzadehkashi2019catalina, ajdari2019cidr} (e.g., SmartSSD~\cite{lee2020smartssd, wang2024ndsearch}). 
While SmartSSD~\cite{lee2020smartssd}  places an FPGA near the SSD, \proposal integrates computation inside the SSD-internal DRAM and controllers. Several works also consider other storage technologies like HDDs~\cite{riedel-computer-2001,riedel-vldb-1998,keeton-sigmod-1998, cho2013active} for computation. 
\revC{None of these works leverages SSD's computational capabilities and enhances them to accelerate \RSGA.}
%to achieve a \revB{highly-efficient} system.

\vspace{-1mm}

\section{Conclusion}
 %We propose \proposal, the first in-storage processing architecture that combines heterogeneous resources and hardware acceleration within SSDs to reduce both data movement and computation overheads of \RSGA read mapping.
%\revB{By implementing a hardware/software co-design, \proposal (1) minimizes hardware demands by integrating early signal quantization and filtering of reads with low informative content, and (2) accelerates \RSGA stages using Processing-Near-Memory and Processing-Using-Memory techniques within the SSD-internal DRAM.
%Our evaluation shows that \proposal outperforms the best prior hardware-accelerated work by 2.9$\times$ on average across five datasets while reducing the energy consumption by 72$\times$, demonstrating the effectiveness of tightly integrating processing into the storage system.}

\revC{We propose \proposal, the first in-storage processing architecture that enables multiple Processing-In-Memory paradigms within the SSD to reduce both data movement and computation overheads of \RSGA read mapping. \proposal (1) proposes targeted
software modifications, such as early signal quantization and read filtering, to minimize hardware resources while maintaining accuracy, and (2) provides near-memory computation units within the SSD for accelerating computational steps of the \RSGA pipeline. MARS improves performance over software- and hardware-accelerated state-of-the-art read mapping pipelines by a factor of 93× and 40× while reducing their energy consumption by 427× and 72× on
average across five real-world dataset.}

\vspace{-1mm}

\section*{\gfcrii{Acknowledgments}}

\gfcrii{We thank the anonymous reviewers of ICS 2025, \revB{ISCA 2025 and HPCA 2025} for their feedback. We thank the SAFARI group members for the feedback and stimulating intellectual environment they provide. We
acknowledge the generous gifts from our industrial partners including Google, Huawei, Intel, and Microsoft. 
This work is
supported in part by the ETH Future Computing Laboratory
(EFCL), Huawei ZRC Storage Team, Semiconductor Research
Corporation, AI Chip Center for Emerging Smart Systems
(ACCESS), sponsored by InnoHK funding, Hong Kong SAR,
and European Union’s Horizon programme for research and
innovation [101047160 - BioPIM].}

\balance
{
  \bstctlcite{IEEEexample:BSTcontrol}
  \let\OLDthebibliography\thebibliography
  \renewcommand\thebibliography[1]{
    \OLDthebibliography{#1}
    \setlength{\parskip}{0pt}
    \setlength{\itemsep}{0pt}
  }
  \bibliographystyle{IEEEtran}
  \bibliography{refs}
}
    
\end{document}